\documentclass{emulateapj}

\begin{document}

\newcommand{\mirlum}{L_{\rm 8}}
\newcommand{\ebmv}{E(B-V)}
\newcommand{\lha}{L(H\alpha)}
\newcommand{\lir}{L_{\rm IR}}
\newcommand{\lbol}{L_{\rm bol}}
\newcommand{\luv}{L_{\rm UV}}
\newcommand{\rs}{{\cal R}}
\newcommand{\ugr}{U_{\rm n}G\rs}
\newcommand{\ks}{K_{\rm s}}
\newcommand{\gmr}{G-\rs}

\title{Dust Obscuration and Metallicity at High Redshift: New
  Inferences from UV, H$\alpha$, and $8$~$\mu$\MakeLowercase{m}
  Observations of $\MakeLowercase{z}\sim 2$ Star-Forming
  Galaxies\altaffilmark{1}} \author{\sc Naveen
  A. Reddy\altaffilmark{2,3}, Dawn K. Erb\altaffilmark{4,5}, Max
  Pettini\altaffilmark{6,7}, Charles C. Steidel\altaffilmark{8}, \&
  Alice E. Shapley\altaffilmark{9}}

\altaffiltext{1}{Based, in part, on data obtained at the W.M. Keck
Observatory, which is operated as a scientific partnership among the
California Institute of Technology, the University of California, and
NASA, and was made possible by the generous financial support of the
W.M. Keck Foundation.}

\altaffiltext{2}{National Optical Astronomy Observatory, 950 N Cherry 
Ave, Tucson, AZ 85719}
\altaffiltext{3}{Hubble Fellow}
\altaffiltext{4}{University of California at Santa Barbara, Santa Barbara, 
CA 93106}
\altaffiltext{5}{Spitzer Fellow}
\altaffiltext{6}{Institute of Astronomy, Madingley Road, Cambridge
CB3 OHA, UK}
\altaffiltext{7}{International Centre for Radio Astronomy Research, University
of Western Australia, 35 Stirling Highway, Crawley, WA 6009, Australia}
\altaffiltext{8}{California Institute of Technology, MS 105--24, Pasadena,
CA 91125}
\altaffiltext{9}{University of California at Los Angeles, 430 Portola Plaza,
Box 951547, Los Angeles, CA 90095}

\slugcomment{DRAFT: \today}

\begin{abstract}

We use a sample of 90 spectroscopically-confirmed Lyman Break Galaxies
with H$\alpha$ measurements and {\em Spitzer} MIPS 24~$\mu$m
observations to constrain the relationship between rest-frame 8~$\mu$m
luminosity ($\mirlum$) and star formation rate (SFR) for L$^{\ast}$
galaxies at $z\sim 2$.  We find a tight correlation with $0.24$~dex
scatter between $\mirlum$ and H$\alpha$ luminosity/SFR for $z\sim 2$
galaxies with $10^{10}\la \lir\la 10^{12}$~L$_{\odot}$.  Employing
this relationship with a larger sample of 392 galaxies with
spectroscopic redshifts, we find that the UV slope $\beta$ can be used
to recover the dust attenuation of the vast majority of moderately
luminous L$^{\ast}$ galaxies at $z\sim 2$ to within a 0.4~dex scatter
using the local correlation.  Separately, young galaxies with ages
$\la 100$~Myr appear to be less dusty than their UV slopes would imply
based on the local trend and may follow an extinction curve that is
steeper than what is typically assumed.  Consequently, very young
galaxies at high redshift may be significantly less dusty than
inferred previously.  Our results provide the first direct evidence,
{\em independent of the UV slope}, for a correlation between UV and
bolometric luminosity ($\lbol$) at high redshift, in the sense the
UV-faint galaxies are also on average less infrared and less
bolometrically-luminous than their UV-bright counterparts.  The
$\lbol$-$\luv$ relation indicates that as the SFR increases, $\luv$
turns over (or ``saturates'') around the value of $L^{\ast}$ at $z\sim
2$, implying that dust obscuration may be largely responsible for
modulating the bright-end of the UV luminosity function.  Finally,
dust attenuation is found to correlate with oxygen abundance at $z\sim
2$.  Accounting for systematic differences in local and high-redshift
metallicity determinations, we find that $L^{\ast}$ galaxies at $z\sim
2$, while at least an order of magnitude more bolometrically-luminous,
exhibit ratios of metals-to-dust that are similar to those of local
starbursts.  This result is expected if high-redshift galaxies are
forming their stars in a less metal-rich environment compared to local
galaxies of the same luminosity, thus naturally leading to a redshift
evolution in both the bolometric luminosity - metallicity and
bolometric luminosity - obscuration relations.

\end{abstract}

\keywords{galaxies: abundances --- galaxies: evolution --- galaxies: high-redshift ---
infrared: galaxies --- ISM: dust, extinction}

\section{INTRODUCTION}
\label{sec:intro}

Quantifying the total energetics and the time evolution of the dust
properties of high-redshift galaxies requires an understanding of the
extent to which the locally calibrated relations between mid-infrared
(rest-frame $8$~$\mu$m) and bolometric luminosity apply at high
redshift.  The {\it Spitzer} Space Telescope allows for a direct
measure of the mid-IR dust emission from typical ($L^{\ast}$) galaxies
at $z>1.5$ and has motivated a number of investigations of the
correlation between mid-IR and bolometric luminosity.  The mid-IR
emission from $5-8.5$~$\mu$m (rest-frame) arises from the stochastic
UV photo-heating of small dust grains and polycyclic aromatic
hydrocarbons (PAHs; e.g., \citealt{puget89, tielens99}).  As such,
this emission is found to correlate with the UV radiation from OB
stars and hence the global star formation rate in nearby galaxies
(e.g., \citealt{forster04b, roussel01}), albeit with some variation
depending on metallicity and ionizing hardness (e.g.,
\citealt{engelbracht05, hogg05, helou01, normand95}.  More recently,
\citet{kennicutt09} demonstrate for the local {\em SINGSs} sample of
galaxies the general agreement between Balmer-decrement dust-corrected
H$\alpha$ star formation rates and those derived by combining tracers
of obscured star formation, including IR and $8$~$\mu$m luminosity,
and tracers of unobscured star formation (observed H$\alpha$
luminosity).

At high redshift, several studies suggest that on average the ratios
of mid to total infrared luminosity for luminous $24$~$\mu$m-selected
star-forming galaxies at $z\sim 2$ are larger than those found for
local galaxies with similar bolometric luminosities \citep{papovich07,
  rigby08}.  On the other hand, X-ray and radio stacking analyses
suggest that the correlation between mid-IR and infrared luminosity
for the less luminous but more typical galaxies at $z\sim 2$ (i.e.,
those with luminosities comparable to $L^{\ast}_{\rm UV}$ or
$L^{\ast}_{\rm bol}$ at $z\sim 2$; \citealt{reddy08, reddy09}) is
consistent with the local relations \citep{reddy06a, reddy04}.  Taken
together, these results imply a luminosity-dependence in the
correlation between mid-IR and bolometric luminosity.

While progress in connecting the dust emission of high-redshift
galaxies to their bolometric luminosities can be informative, most
$z>3$ galaxies are selected via the efficient UV-dropout technique
\citep{steidel95} and tend to be too faint and at too high redshift to
be detected directly via their dust emission.  Thus, it has become
common to rely on UV-based inferences of the dust attenuation and
bolometric star formation rates of $z>3$ galaxies that require some
assumption of how the rest-frame UV slope varies with extinction.  The
highest redshift where the correlation between UV-slope and dust
extinction has been tested for large numbers of
spectroscopically-confirmed galaxies is at $z\sim 2$.  The results
suggest that the local relation between UV-slope and extinction
remains valid at $z\sim 2$ for galaxies with moderate bolometric
luminosities similar to those of luminous infrared galaxies, or LIRGs
(\citealt{reddy06a}; hereafter R06).

Access to a third tracer of star formation, independent of the UV and
IR emission, is required to establish a more secure footing for (1)
the scaling between mid-IR emission and star formation rate and (2)
the dependence of UV-slope on dust extinction.  The most direct probe
of young massive star formation is from HII recombination nebular
emission.  In particular, the H$\alpha$ line has been used
traditionally as a star formation rate indicator given its
accessibility in the optical window and the fact that it traces star
formation on very short timescales ($\sim 10$~Myr) and is less
sensitive to extinction than UV emission (e.g., \citealt{kennicutt98,
  brinchmann04}).  Fortunately, H$\alpha$ is still accessible with
ground-based near-IR spectrographs for galaxies at $z\sim 2$.  Using a
sample of $114$ UV-selected galaxies at $z\sim 2$ with H$\alpha$
spectroscopy, \citet{erb06c} demonstrate that the H$\alpha$ inferred
star formation rates agree well with those obtained from the UV after
correcting both the H$\alpha$ line and the UV continuum magnitude for
extinction, and assuming that $\ebmv_{\rm stellar} \approx \ebmv_{\rm
  nebular}$.  In this paper, we use this H$\alpha$ spectroscopic
sample as the basis for quantifying independently the scaling between
mid-IR emission and star formation rate for moderately-luminous
galaxies at $z\sim 2$.  This scaling relation is then used to infer
bolometric luminosities and dust obscuration and determine how the UV
slope varies with extinction for typical high-redshift galaxies.

We begin by discussing in \S~\ref{sec:selection} the rest-UV
selection, followup optical and near-IR spectroscopy, and the MIPS
observations, data reduction and photometry.  The stellar population
modeling of galaxies in our final samples is discussed in
\S~\ref{sec:sedfit}.  We then present the correlations between
H$\alpha$, X-ray, and $8$~$\mu$m luminosity (and the connection
between these quantities and the star formation rate) and comparison
with other local and high-redshift relations (\S~\ref{sec:ha24}).  The
variation of UV slope with dust attenuation is addressed in
\S~\ref{sec:meurer}.  The correlation between bolometric luminosity
and dust attenuation and the implications of this relationship for the
observed UV luminosity are presented in \S~\ref{sec:bol}.  Finally, in
\S~\ref{sec:metals} we present the results of a comparison between
extinction and gas-phase metallicities of star-forming galaxies at
$z\sim 2$ and demonstrate a close relationship between the two.  We
assume a \citet{chabrier03} IMF unless stated otherwise and adopt a
cosmology with $H_{0}=70$~km~s$^{-1}$~Mpc$^{-1}$,
$\Omega_{\Lambda}=0.7$, and $\Omega_{\rm m}=0.3$.

\section{SAMPLE}
\label{sec:selection}

\subsection{Optical and H$\alpha$ Observations and Spectroscopy}

The photometry and spectroscopic followup for the Lyman Break Galaxy
(LBG) survey at redshifts $1.4\la z\la 3.4$ are described in
\citet{steidel03, steidel04, adelberger04}.  Briefly, galaxies were
photometrically-selected using either the ``BM,'' ``BX,'' or LBG
rest-UV color criteria.  A subset of $\rs<25.5$ candidates that were
spectroscopically-confirmed with Keck/LRIS observations
\citep{steidel04} were targeted with NIRSPEC H or K-band spectroscopy
to trace H$\alpha$ \citep{erb06b}.  In addition to the longslit
H$\alpha$ spectroscopy, we targeted the $z=2.300$ redshift overdensity
in the HS1700+64 (``Q1700'') field \citep{steidel05} with narrowband
Br$\gamma$ ($2.17$~$\mu$m/$0.04$~$\mu$m) observations that trace
H$\alpha$ at that redshift.  These images, along with $\ks$-band
observations, were obtained with the Wide-Field Infrared Camera (WIRC)
on the Palomar $5$~m telescope.  Details on the Br$\gamma$ imaging
reductions, continuum subtraction, and H$\alpha$ flux determinations
will be provided in a forthcoming paper (Steidel et~al., in prep.).
Briefly, we calibrated the Br$\gamma$ image by adjusting its zeropoint
until the difference between $\ks$ and Br$\gamma$ magnitude was zero
for objects of similar faintness in $\ks$.  Colors were measured in
matched isophotal apertures with detection done at $\ks$-band.  We
then measured the narrowband fluxes and subtracted the continuum based
on the Br$\gamma$-$\ks$ colors and knowledge of the Br$\gamma$ filter
bandwidth and shape.  Photometric and zeropoint uncertainties result
in an $\simeq 20\%$ error in the derived line fluxes.

This procedure resulted in 21 objects with $>5$~$\sigma$ flux limits
in H$\alpha$, ranging in flux from $\sim 5\times 10^{-17}$ to $\sim
3\times 10^{-16}$~ergs~s$^{-1}$.  Eleven of these 21 objects also have
longslit H$\alpha$ observations.  Combining the longslit and
narrowband H$\alpha$ samples yields a total number of 116 galaxies in
6 fields that have $24$~$\mu$m coverage and are spectroscopically
confirmed to lie at $1.5\le z\le 2.6$, where the $24$~$\mu$m data are
sensitive to the rest-frame $8$~$\mu$m flux (\S~\ref{sec:mips}).

\subsection{Correction for Slit Loss}
\label{sec:slitloss}

The primary uncertainty in the spectroscopically-derived H$\alpha$
fluxes is slit loss.  The overlap of 11 objects between the H$\alpha$
longslit and narrowband samples in the Q1700 field allows us to
quantify this source of error.  To assess the degree of [NII]
contamination of the narrowband fluxes, we constructed a composite
spectrum of the 11 objects with both longslit and narrowband
observations.  From this composite, we determine a mean ratio of the
[NII]-to-H$\alpha$ flux of $\approx 0.09$ for these 11 objects.  At
the redshift of these galaxies ($z=2.300$), the Br$\gamma$
transmission curve will suppress the [NII] line by a factor of 0.88
relative to H$\alpha$.  Therefore, we expect the average [NII] line
contamination of the narrowband flux to be $\approx 8\%$, translating
to a $0.03$~dex correction in log flux.  After taking into account
this $8\%$ correction, the ratio of the narrowband H$\alpha$ flux to
the spectroscopic H$\alpha$ flux for these 11 objects varies from
$1.14$ to $4.21$, with a median value of $1.87$ and a $1$~$\sigma$
dispersion around this median value of $\sim 0.8$ (i.e., a scatter of
$\approx 40\%$).  This median value is similar to the average slit
loss correction factor of $2.0$ suggested by \citet{erb06c}.  For
galaxies with near-IR spectroscopic observations only, we correct the
H$\alpha$ flux for slit losses by multiplying by the median value
found above, $1.87$.

The sample of 11 galaxies with both longslit and narrowband
observations have an H$\alpha$ flux distribution that is
undistinguished from the distribution for the 90 galaxies that are
used to measure the relation between 8 micron luminosity ($\mirlum =
\nu L_{\nu}(8\mu m)$) and $\lha$.  Over this flux range, we find no
dependence of the slit loss correction factor with the spectroscopic
flux.  The absence of any spectroscopic flux-dependent systematic in
the slit loss correction, combined with the expectation that the
amount of H$\alpha$ flux lost in the spectroscopic observations
depends sensitively on the relative placement and position angle of
the slit on the galaxy, leads us to the conclusion that the slit loss
correction must be dominated by random error.

\subsection{MIPS Observations}
\label{sec:mips}

There are 6 fields in the LBG survey that have {\it Spitzer} MIPS
$24$~$\mu$m coverage, including the publicly available data in the
GOODS-N and Westphal fields (Dickinson et~al., in prep.; Chary et~al.,
in prep.).  We obtained observations in 4 additional LBG fields:
Q1623, Q1700, Q2343, and Q1549 (see \citealt{reddy09}).  Observations
in these 4 fields consisted of 30~sec frame times with 2 mapping
cycles, where each cycle consists of a $3\times 3$ dither pattern with
a step size of $1/8$ the size of the MIPS array.  The total exposure
time in 3 of the 4 fields is $\approx 11.4$~hours, sufficient to
detect $\approx 15$~$\mu$Jy sources at $>3$~$\sigma$ (in the absence
of confusion) within the inner $4\arcmin$ of each field.  This is
comparable to the depth of the $24$~$\mu$m data in the GOODS-N and
Westphal fields.  The Q1700 MIPS data consist of $\approx 20$~hrs of
observations and reach a $3$~$\sigma$ depth of $\approx 10$~$\mu$Jy in
the absence of confusion.

The $24$~$\mu$m images are processed using a procedure similar to that
described in R06.  Briefly, the data are flatfielded and then combined
using MOPEX \citep{makovoz05}.  Photometry is extracted using PSF
photometry with priors determined by the locations of objects detected
in the IRAC images (IRAC data exist in all 6 fields).  There are three
sources of uncertainty in the $24$~$\mu$m fluxes.  The first is
background noise, which we determine by fitting PSFs simultaneously to
random positions around the target of interest and any other known
sources.  The dispersion in measurements of the background level
obtained in this way is similar to the dispersion we would estimate by
comparing the simulated and recovered fluxes of artificial objects
added to the $24$~$\mu$m images.  The second source of uncertainty is
Poisson error and is generally negligible compared to the background
dispersion.  The third source of uncertainty is confusion, which is
reflected by the degeneracy between the PSF fit to the object of
interest and the fits to any nearby sources.  The uncertainty in the
fit is computed simply from the covariance matrix for the $N$ number
of objects that are being fit simultaneously.  The total (normalized)
covariance in the fit ($\sigma_{\rm cov}$) is a number between 0 and
1, where the former indicates no covariance with nearby objects and
the latter indicates total covariance (e.g., such as the covariance
between the fit of an object and itself).  For most galaxies in the
sample of $116$, $\sigma_{\rm cov}<<1$.  There are 17 out of 116
galaxies that have $\sigma_{\rm cov}>0.3$.  We consider these sources
to be confused and remove them from our sample; this results in 99
galaxies with secure PSF fits.

\subsection{Identification of AGN}
\label{sec:agn}

Because we are interested primarily in the star-forming galaxies, we
have removed AGN from the sample based on (1) the presence of
prominent emission lines including Ly$\alpha$, CIV, and NV, or (2) an
SED that behaves as a power-law between rest-frame $\sim 2$~$\mu$m and
$8$~$\mu$m (\S~\ref{sec:sedfit}).  Among the 99 unconfused galaxies in
our sample, 9 are identified as AGN.  Roughly half of these AGN also
exhibited high ratios of [NII]/H$\alpha \ga 0.5$.  The fraction of AGN
($\approx 9\%$) is somewhat larger than the $\sim 5\%$ found among
typical LBGs to ${\cal R}=25.5$ \citep{reddy06b}; the H$\alpha$
spectroscopic sample is biased towards more K-bright galaxies relative
to the typical LBG \citep{erb06c}, and the frequency of AGN increases
rapidly with increasing {\em K}-band brightness at $z\sim 2$
\citep{reddy05a}.

\subsection{Final Samples}
\label{sec:finalsample}

The sample used to measure the relationship between mid-IR luminosity
and star formation rate consists of 90 galaxies in the H$\alpha$
sample (including spectroscopic and narrowband selected objects) that
satisfy the following criteria: (1) $1.5\le z_{\rm spec}\le 2.6$, (2)
$\sigma_{\rm cov}<0.3$, and (3) do not show AGN signatures based on
optical emission lines or a power-law SED.  Of these 90 galaxies, 29
are undetected at $24$~$\mu$m to $3$~$\sigma$.  We use a larger sample
of 158 UV-selected galaxies in the GOODS-N field that satisfy the
three criteria specified above and have corresponding X-ray data that
allow us to perform an X-ray stacking analysis (\S~\ref{sec:stack}).
Finally, the relationship between dustiness and UV spectral slope is
investigated with a sample of 392 galaxies (from 5 of the 6 LBG fields
with MIPS data) that satisfy the three criteria above and have had
their photometry modeled using a stellar population analysis
(\S~\ref{sec:sedfit}).

\section{Stellar Population Modeling}
\label{sec:sedfit}

To provide a broader context for interpreting the extinction
properties, we have derived ages and stellar masses by modeling the
broadband photometry of galaxies in our sample.  All of the galaxies
with MIPS data have been imaged with IRAC, enabling us to use the full
rest-frame UV through near-IR photometry to fit for their stellar
populations.  Previous efforts in this regard are described in
\citet{shapley05}, \citet{erb06b}, and \citet{reddy06b}.  Here, we
update these results using the latest models of Charlot \& Bruzual (in
prep; CB08) that include the \citet{marigo07} prescription for the
thermally-pulsating AGB (TP-AGB) evolution of low- and
intermediate-mass stars.  The effect of this evolution on the inferred
ages and stellar masses is then determined by comparing with our
previous results that employed the \citet{bruzual03} (BC03) models.

\begin{figure*}[!t]
\plottwo{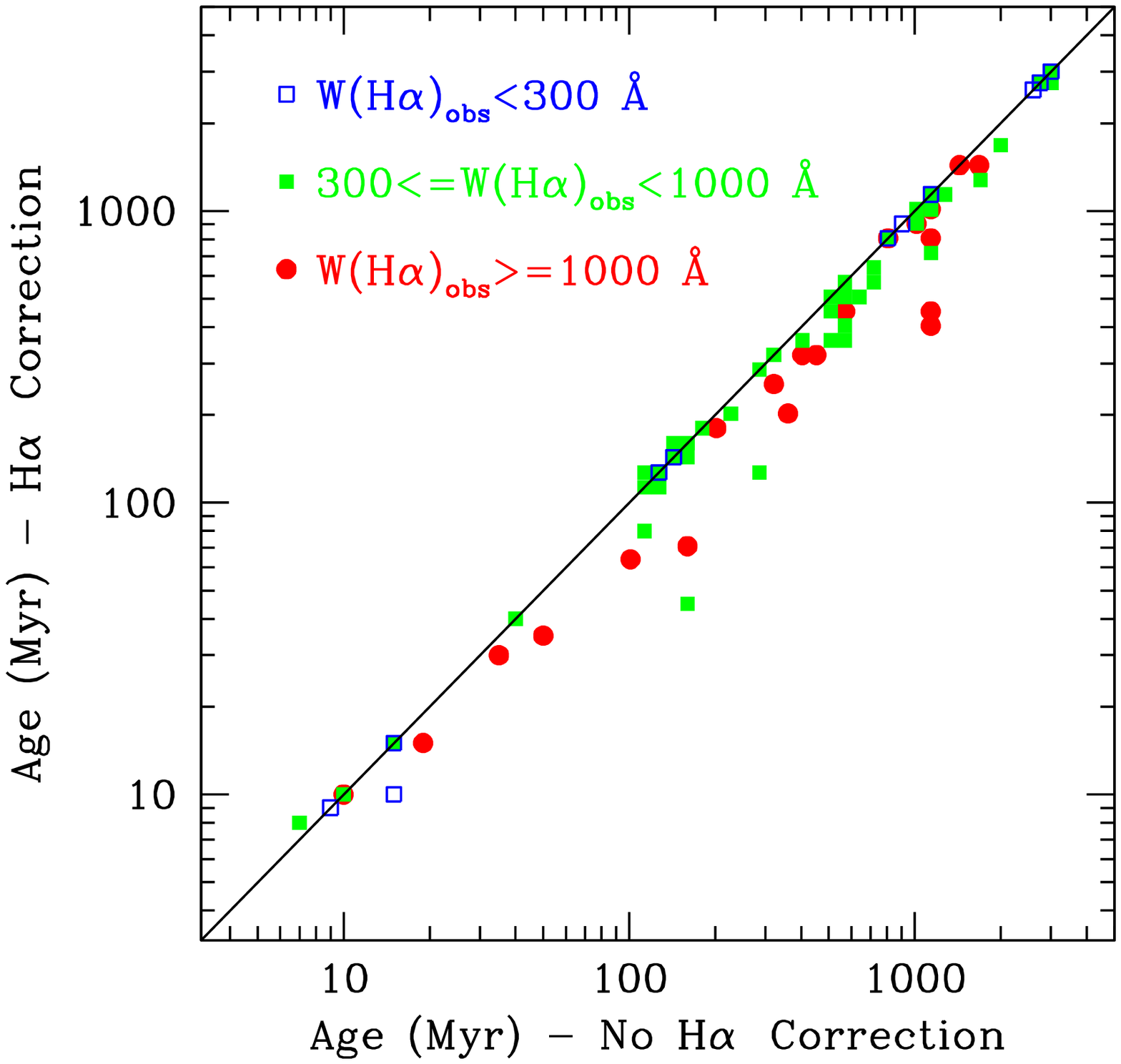}{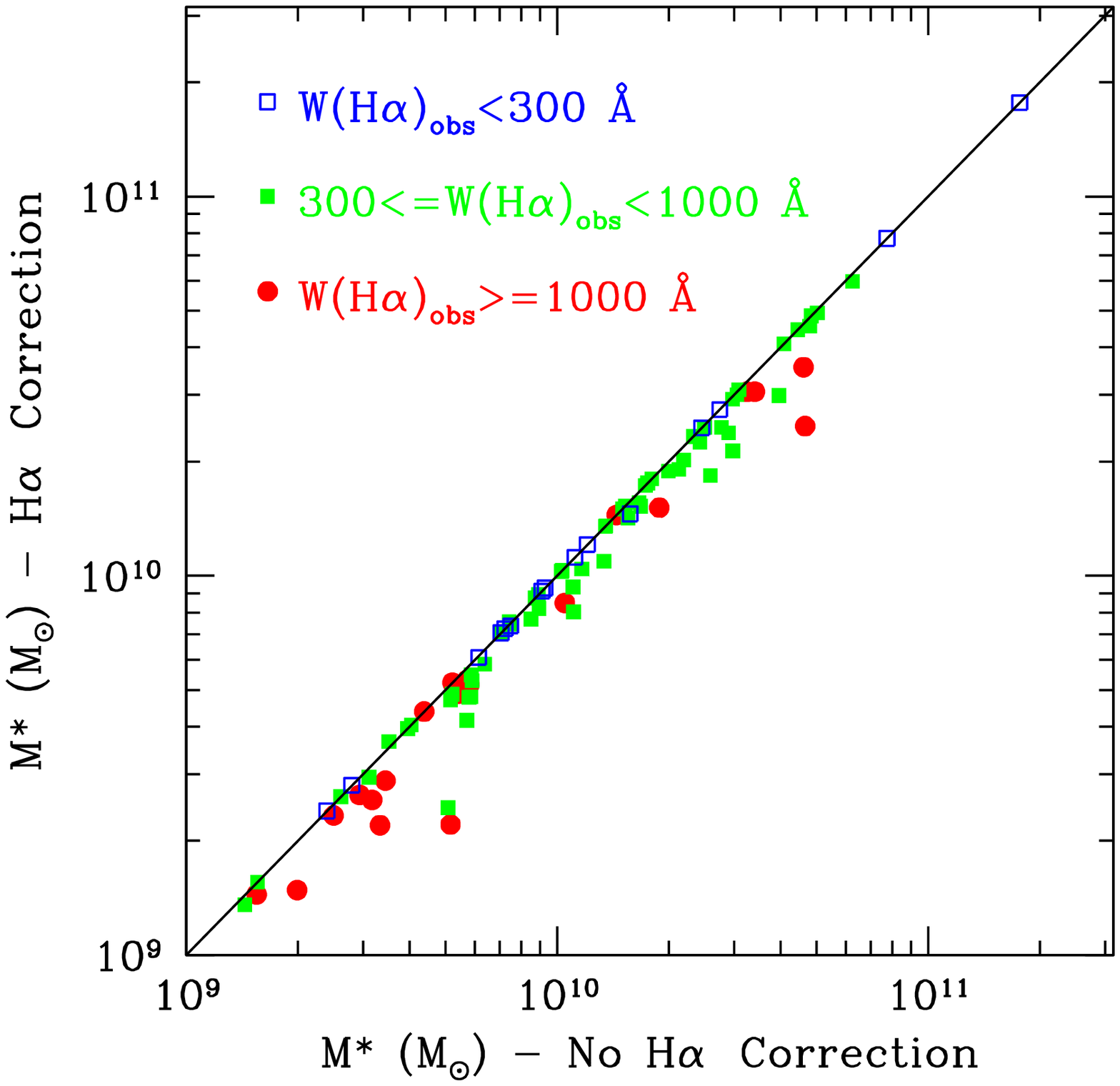}
\caption{Comparison of the best-fit ages ({\em left}) and stellar
  masses ({\em right}) of 95 galaxies with H$\alpha$ measurements
  before and after correcting the $\ks$-band flux for H$\alpha$ line
  contamination.  Points are color-coded according to observed
  H$\alpha$ equivalent width.}
\label{fig:sedcomp}
\end{figure*}

\subsection{Modeling Procedure}

For each galaxy, we considered a constant star-formation (CSF) model
and exponentially-declining star formation histories with
characteristic time-scales $\tau =$ 10, 20, 50, 100, 200, 500, 1000,
2000, and 5000 Myr.  We further considered a range of ages spaced
roughly logarithmically between 1 and 5000 Myr, excluding ages older
than the age of the universe at the redshift of each galaxy.  In
addition, allowed ages were restricted to be longer than the dynamical
time-scale of $\simeq 70$~Myr, as inferred from velocity dispersion
and size measurements of $z\sim 2$ LBGs \citep{erb06c, law07}.
Reddening is taken into account by employing the Calzetti extinction
curve and allowing $\ebmv$ to range between 0.0$-$0.6.  In
S~\ref{sec:ebmvyoung} we also consider the effects of adopting the SMC
extinction curve on the model results for the youngest galaxies in our
sample.  The model SED at each $\tau$ and age combination is reddened,
redshifted, and attenuated blueward of rest-frame $1216$~\AA\, for the
opacity of the IGM using the \citet{madau95} prescription.  The
best-fit normalization of this model is determined by minimizing its
$\chi^2$ with respect to the observed $\ugr+J\ks$+IRAC
(3.6$-$8.0~$\mu$m) photometry.  This normalization then determines the
star formation rate and stellar mass.  The model (and normalization)
that gives the lowest $\chi^2$ is taken to be the best-fit SED.  We
note the inherent degeneracy in this type of modeling.  Typically
there are several best-fit models that may adequately describe the
observed photometry, even when the redshift is fixed to the
spectroscopic value, though there is generally less variation in
stellar mass than in the other parameters ($\tau$, age, $\ebmv$) among
these best-fit models \citep{shapley05, shapley01, papovich01,
  sawicki98}.

In the subsequent analysis, we have adopted the best-fit parameters
obtained with a CSF model for several reasons.  This model generally
yields $\chi^{2}$ values similar to those obtained when $\tau$ is
allowed to vary freely.  Additionally, some of the more extreme star
formation histories are ruled out because the youngest galaxies in our
sample cannot realistically have ages much less than the dynamical
time-scale of $\simeq 70$~Myr.  Such extreme models are also unlikely
based on the presence of O star and Wolf-Rayet stellar features in the
composite UV spectra of $z\sim 2$ galaxies, irrespective of age (e.g.,
\citealt{shapley05}).  For simplicity, we do not consider more complex
multi-component or stochastic star formation histories as the data
would not allow for discrimination between them as compared to simpler
star formation histories.  Finally, we do not use the SFRs from the
SED fitting in the subsequent analysis, opting instead to use the
direct tracers examined here, including UV, H$\alpha$, X-ray, and
$8$~$\mu$m luminosity.

\subsection{Effect of H$\alpha$ Emission on the SEDs}

The parameters of most interest are the age and stellar mass, both of
which are determined primarily by the strength of the Balmer and
4000~\AA\, breaks and the stellar continuum flux as traced by the
near-IR and IRAC photometry at $z\sim 2$.  Because the H$\alpha$
emission line falls in the $\ks$-band at $z\sim 2.0-2.6$, it is
prudent to determine what effect, if any, the line has on biasing our
estimates of the ages and stellar masses.  Figure~\ref{fig:sedcomp}
summarizes the comparison of the best-fit ages and stellar masses
(assuming constant star formation) obtained with and without
correcting for the H$\alpha$ contribution to the $\ks$-band flux for
95 galaxies with H$\alpha$ measurements.  Roughly two-thirds of the
sample of 95 galaxies show no change in best-fit age or stellar mass
after correcting the $\ks$-band flux for H$\alpha$ line contamination,
and not surprisingly these galaxies generally have small H$\alpha$
equivalent widths ($W_{\rm obs}\la 300$~\AA).  Even for those galaxies
with larger H$\alpha$ equivalent widths, the mean difference in age
and stellar mass is relatively small compared to the uncertainty in
these parameters as a result of the significant degeneracies inherent
in SED fitting.

The small differences in the ages and stellar masses before and after
correcting for H$\alpha$ emission partly reflect the fact that these
parameters are not solely constrained by the near-IR data: 89 of the
95 galaxies also have higher $S/N$ IRAC data that are unaffected by
line contamination at these redshifts.  Also note that the age
distribution for galaxies with the largest H$\alpha$ equivalent widths
($W_{\rm obs}\ga 1000$~\AA), including many of those selected from the
narrowband data of the Q1700 field (\S~\ref{sec:selection}), is not
significantly different from that of galaxies with smaller equivalent
widths.  The lack of a significant trend between H$\alpha$ equivalent
width and age for this sample has been noted by \citet{erb06c}, and
may be due to the stochasticity of the instantaneous SFR.  In any
case, given that the age and SFR distribution of galaxies with
H$\alpha$ measurements span roughly the entire range probed by
UV-selected galaxies in general \citep{erb06b}, and that the
correction for the H$\alpha$ flux does not significantly alter the
best-fit ages and stellar masses for the majority of the sample, we
are confident in our ability to constrain the SED fits of galaxies at
$z\sim 2.0-2.6$ even in the absence of direct measurements of the
H$\alpha$ line contamination in the $\ks$-band.

\subsection{Comparison between CB08 and BC03}

\begin{figure}[!t]
\plotone{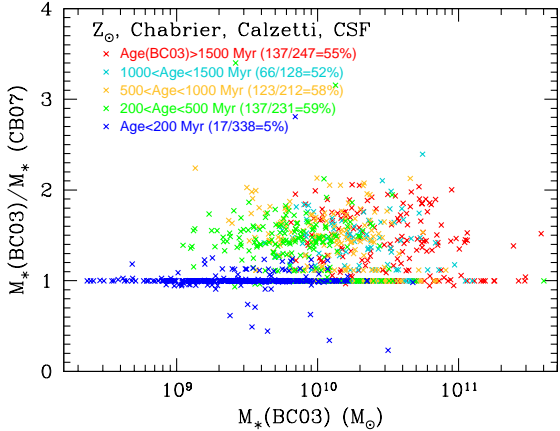}
\caption{Ratio of BC03 to CB08 derived stellar masses as a function of
  BC03 derived stellar mass for 1156 UV-selected galaxies with
  spectroscopic redshifts $1.40<z<3.50$.  Fractions indicate the
  percentage of objects in each age range that have CB08 masses that
  deviate more than $20\%$ from the BC03 mass. }
\label{fig:bc03cb07}
\end{figure}

Figure~\ref{fig:bc03cb07} compares the BC03 and CB08 derived stellar
masses for 1156 UV-selected galaxies with spectroscopic redshifts
$1.40<z<3.50$, color coded by the BC03-derived stellar age.  The
TP-AGB phase will become important $\simeq 200$~Myr after the onset of
star formation.  Anywhere from $50-60\%$ of galaxies older than this
have CB08 derived stellar masses that deviate by more than $20\%$ from
their BC03 derived stellar mass, with a median CB08 stellar mass that
is a factor of $\sim 1.5$ times lower than the BC03 stellar mass.  The
CB08 models also imply ages that are on average $100$~Myr younger than
BC03-derived ages.  We will make use of the CB08 derived ages, stellar
masses, and color excesses in the subsequent analysis.

\section{DEPENDENCE OF $8$~$\mu$\MakeLowercase{m} LUMINOSITY ON STAR FORMATION
RATE}
\label{sec:ha24}

\subsection{Computing the Mid-IR and H$\alpha$ Luminosities}

The $8$~$\mu$m luminosity, $\mirlum$, is computed by {\em
  k}-correcting the $24$~$\mu$m flux using the average mid-IR spectral
shape of local starburst galaxies, as discussed in R06.  All 90
galaxies have {\em spectroscopic} redshifts and therefore are immune
to the additional $0.3$~dex uncertainty introduced by photometric
redshift errors typical of $z\sim 2$ star-forming galaxies, owing to
the complicated mid-IR spectral shape and the redshift-sensitivity of
the {\em k}-correction (R06).

H$\alpha$ luminosities are corrected for extinction based on the
$\ebmv$ color excess as derived from the SED fitting
(\S~\ref{sec:sedfit}).  The average color excess is $\langle
\ebmv\rangle = 0.15 \pm 0.10$, implying an average UV attenuation of a
factor of $\sim 4-5$ \citep{reddy04, reddy06a}.  Similar average dust
obscurations have been obtained based on stacked X-ray analyses of
LBGs at $z\sim 3$ (e.g., \citealt{seibert02, nandra02, reddy04}).
Adopting the conversion $\ebmv_{\rm stellar} = 0.4\ebmv_{\rm nebular}$
advocated by \citet{calzetti00} results in H$\alpha$-inferred SFRs
that notably overpredict those derived from the X-ray and
dust-corrected UV for the sample presented here\citep{erb06c}.
Therefore, the same value of $\ebmv$ is assumed for both the UV
continuum and nebular emission lines.\footnote{\citet{forster09} find
  evidence in their sample of $z\sim 2$ galaxies targeted with {\em
    VLT}/SINFONI observations that the extinction of the nebular
  regions may be larger than that for the stellar continuum, though
  their sample includes galaxies with somewhat larger SFRs and larger
  stellar masses than those typical of galaxies in our sample.  It is
  possible that the nebular reddening may on average be larger for
  galaxies that are forming stars at a higher rate and/or hosting an
  older (less attenuated) stellar population (e.g.,
  \citealt{epinat09}).  While we note that some small amount of
  additional nebular extinction may be allowable by our data after
  correcting for slit losses (given the $\approx 40\%$ random scatter
  in such losses), this systematic difference in attenuation is likely
  to be small relative to the scatter in the slit loss correction and
  the scatter in correlation between dust-corrected H$\alpha$ and
  dust-corrected UV SFRs of $\approx 0.3$~dex.  Given that the
  galaxies (with H$\alpha$ observations) analyzed here are drawn from
  the same sample presented in \citet{erb06c}, and the fact that these
  authors found no significant evidence for a larger obscuration
  towards the nebular regions in these galaxies, we proceed under the
  assumption that $\ebmv_{\rm stellar}=\ebmv_{\rm nebular}$.}  The
\citet{calzetti00} extinction curve then implies a mean H$\alpha$
extinction correction of $1.71\pm 0.40$.  The uncertainty in the
observed H$\alpha$ luminosity is dominated by random error in the slit
loss correction factor and is taken to be $\approx 40\%$
(\S~\ref{sec:slitloss}).

\begin{deluxetable*}{lccccccc}
\tabletypesize{\footnotesize}
\tablewidth{0pc}
\tablecaption{Stacked X-ray and H$\alpha$ and UV Luminosities}
\tablehead{
\colhead{Sample} &
\colhead{N~\tablenotemark{a}} &
\colhead{$F_{\rm 0.5-2.0~keV}$~\tablenotemark{b}} &
\colhead{$L_{\rm 2.0-10.0~keV}$~\tablenotemark{c}} &
\colhead{$L_{\rm H\alpha}^{obs}$~\tablenotemark{d}} &
\colhead{$L_{\rm H\alpha}^{cor}$~\tablenotemark{e}} &
\colhead{$L_{\rm UV}^{obs}$~\tablenotemark{f}} &
\colhead{$L_{\rm UV}^{cor}$~\tablenotemark{g}} \\ \\
\colhead{(1)} &
\colhead{(2)} &
\colhead{(3)} &
\colhead{(4)} &
\colhead{(5)} &
\colhead{(6)} &
\colhead{(7)} &
\colhead{(8)}}
\startdata
H$\alpha$ (All) & 18 &  $6.6\pm2.4$ & $3.4\pm1.3$ & $4.3\pm2.6$ & $7.4\pm4.4$ & $11.5\pm6.3$ & $54.4\pm30.0$ \\
H$\alpha$ (Undet)~\tablenotemark{h} & 8 & $3.3\pm1.5$ & $1.8\pm0.8$ & $3.5\pm2.2$ & $4.9\pm3.1$ & $12.2\pm5.9$ & $32.4\pm15.8$ \\
\\
UV (All) & 158 & $5.1\pm1.8$ & $1.8\pm0.6$ & ... & ... & $7.6\pm4.1$ & $41.0\pm22.3$ \\
UV (Undet)~\tablenotemark{h} & 116 & $2.8\pm1.7$ & $1.0\pm0.6$ & ... & ... & $7.4\pm3.9$ & $29.3\pm15.4$ \\
\\
UV ($0<{\rm SFR}<20$)~\tablenotemark{i} & 58 & $2.4\pm1.5$ & $0.9\pm0.6$ & ... & ... & $5.9\pm2.6$ & $15.7\pm7.0$ \\
UV ($20<{\rm SFR}<40$)~\tablenotemark{i} & 43 & $9.6\pm4.6$ & $3.3\pm1.6$ & ... & ... & $7.9\pm3.3$ & $42.9\pm17.9$ \\
UV ($40<{\rm SFR}<80$)~\tablenotemark{i} & 32 & $12.2\pm5.0$ & $4.6\pm1.9$ & ... & ... & $8.4\pm5.3$ & $81.4\pm31.1$ \\
\\
UV ($0<{\rm SFR}<20$; Undet)~\tablenotemark{h,i} & 56 & $2.0\pm1.7$ & $0.7\pm0.6$ & ... & ... & $5.9\pm2.7$ & $15.7\pm7.1$ \\
UV ($20<{\rm SFR}<40$; Undet)~\tablenotemark{h,i} & 28 & $5.8\pm3.1$ & $2.3\pm1.2$ & ... & ... & $8.7\pm3.5$ & $46.9\pm18.7$ \\
UV ($40<{\rm SFR}<80$; Undet)~\tablenotemark{h,i} & 16 & $8.4\pm4.4$ & $3.9\pm2.1$ & ... & ... & $8.9\pm5.6$ & $78.2\pm29.5$ \\
\enddata
\tablenotetext{a}{Number of galaxies in sample.}
\tablenotetext{b}{Average soft-band X-ray flux, in units of
$10^{-18}$~ergs~s$^{-1}$~cm$^{-2}$.}
\tablenotetext{c}{Average X-ray luminosity in units of
$10^{41}$~ergs~s$^{-1}$, assuming an
X-ray photon index of $\Gamma = 2.0$ and Galactic absorption 
column density of $1.6\times 10^{20}$~cm$^{-2}$ \citep{stark92}.}
\tablenotetext{d}{Observed H$\alpha$ luminosity in units of
$10^{42}$~ergs~s$^{-1}$, uncorrected for dust.}
\tablenotetext{e}{Dust-corrected H$\alpha$ luminosity in units
of $10^{42}$~ergs~s$^{-1}$.}
\tablenotetext{f}{Observed UV luminosity in units of 
$10^{28}$~ergs~s$^{-1}$~Hz$^{-1}$, uncorrected for dust.}
\tablenotetext{g}{Dust-corrected UV luminosity in units
of $10^{28}$~ergs~s$^{-1}$~Hz$^{-1}$.}
\tablenotetext{h}{Stack of galaxies not detected at $24$~$\mu$m.}
\tablenotetext{i}{Stack of galaxies with dust-corrected UV-inferred
SFRs in M$_{\odot}$~yr$^{-1}$ as specified, assuming a Chabrier IMF.}
\label{tab:lum}
\end{deluxetable*}

\begin{deluxetable*}{lccccccccc}
\tabletypesize{\footnotesize}
\tablewidth{0pc}
\tablecaption{Star Formation Rates and Attenuation from X-ray, H$\alpha$, and UV}
\tablehead{
\colhead{Sample~\tablenotemark{a}} &
\colhead{SFR$_{\rm X}$~\tablenotemark{b}} &
\colhead{SFR$_{\rm H\alpha}^{obs}$~\tablenotemark{b}} &
\colhead{SFR$_{\rm H\alpha}^{cor}$~\tablenotemark{b}} &
\colhead{{\sc A}$_{\rm H\alpha}^{\rm \ebmv}$~\tablenotemark{c}} &
\colhead{{\sc A}$_{\rm H\alpha}^{\rm X-ray}$~\tablenotemark{d}} &
\colhead{SFR$_{\rm UV}^{obs}$~\tablenotemark{b}} &
\colhead{SFR$_{\rm UV}^{cor}$~\tablenotemark{b}} &
\colhead{{\sc A}$_{\rm UV}^{\rm \ebmv}$~\tablenotemark{c}} &
\colhead{{\sc A}$_{\rm UV}^{\rm X-ray}$~\tablenotemark{d}} \\ \\
\colhead{(1)} &
\colhead{(2)} &
\colhead{(3)} &
\colhead{(4)} &
\colhead{(5)} &
\colhead{(6)} &
\colhead{(7)} &
\colhead{(8)} &
\colhead{(9)} &
\colhead{(10)}}
\startdata
H$\alpha$ (All) & $37\pm14$ & $19\pm11$ & $32\pm19$ & $\sim1.7$ & $\sim1.9$ & $9\pm5$ & $42\pm23$ & $\sim4.7$ & $\sim4.1$ \\
H$\alpha$ (Undet) & $20\pm9$ & $15\pm10$ & $22\pm13$ & $\sim1.5$ & $\sim1.3$ & $9\pm5$ & $25\pm12$ & $\sim2.8$ & $\sim2.2$ \\
\\
UV (All) & $20\pm7$ & ... & ... & ... & ... & $6\pm3$ & $32\pm17$ & $\sim5.3$ & $\sim3.3$ \\
UV (Undet) & $12\pm7$ & ... & ... & ... & ... & $6\pm3$ & $23\pm12$ & $\sim3.8$ & $\sim2.0$ \\
\\
UV ($0<{\rm SFR}<20$) & $10\pm6$ & ... & ... & ... & ... & $5\pm2$ & $12\pm5$ & $\sim2.4$ & $\sim2.0$ \\
UV ($20<{\rm SFR}<40$) & $37\pm18$ & ... & ... & ... & ... & $6\pm3$ & $33\pm14$ & $\sim5.5$ & $\sim 6.2$ \\
UV ($40<{\rm SFR}<80$) & $51\pm21$ & ... & ... & ... & ... & $7\pm4$ & $63\pm24$ & $\sim9.0$ & $\sim 7.3$ \\
\\
UV ($0<{\rm SFR}<20$; Undet) & $8\pm7$ & ... & ... & ... & ... & $5\pm2$ & $12\pm6$ & $\sim 2.4$ & $\sim1.6$ \\
UV ($20<{\rm SFR}<40$; Undet) & $25\pm13$ & ... & ... & ... & ... & $7\pm3$ & $36\pm15$ & $\sim5.1$ & $\sim3.6$ \\
UV ($40<{\rm SFR}<80$; Undet) & $44\pm23$ & ... & ... & ... & ... & $7\pm4$ & $61\pm23$ & $\sim8.7$ & $\sim6.3$ \\
\enddata
\tablenotetext{a}{Samples are the same as in Table~\ref{tab:lum}.}
\tablenotetext{b}{In M$_{\odot}$~yr$^{-1}$, assuming a Chabrier IMF, based on the conversion
relations of \citet{ranalli03} and \citet{kennicutt98}.}
\tablenotetext{c}{Attenuation factor corresponding to $\ebmv$ assuming the \citet{calzetti00} relation.}
\tablenotetext{d}{Attenuation factor from the ratio of the X-ray-inferred SFR to the observed H$\alpha$
and UV SFRs.} 
\label{tab:sfr}
\end{deluxetable*}

\subsection{X-ray Stacking Analysis}
\label{sec:stack}

\begin{figure}[!t]
\plotone{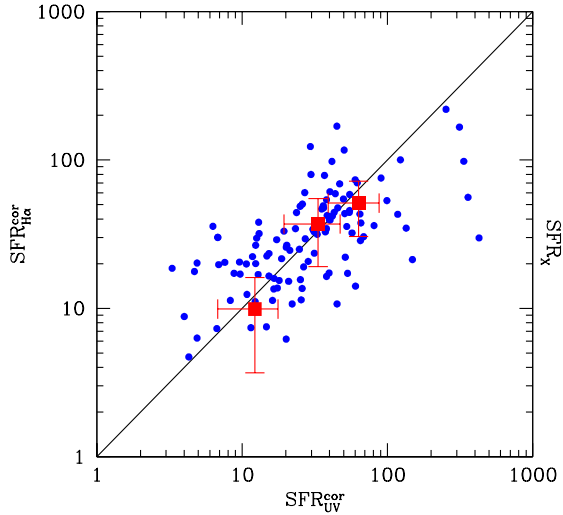}
\caption{Dust-corrected H$\alpha$ SFR versus UV SFR ({\em blue
    symbols}; data from \citealt{erb06c}) compared to X-ray determined
  SFRs ({\em red symbols}).}
\label{fig:ha_uv}
\end{figure}

Recall that our main goal is to determine how mid-IR luminosity scales
with SFR by cross-correlating the {\em Spitzer} data with
dust-corrected H$\alpha$ luminosity.  We have two independent measures
of how well the dust-corrected H$\alpha$ luminosity scales with SFR.
For instance, \citet{erb06c} and \citet{reddy04} demonstrate the
agreement between dust-corrected UV, dust-corrected H$\alpha$, and
stacked X-ray estimates of the SFRs of the same types of galaxies at
$z\sim 2$ that are examined in this study.  To provide better
constraints, and to take advantage of the new spectroscopy done after
these initial studies, we revisited the use of X-ray emission as a
proxy for SFR in our larger UV spectroscopic sample.  To accomplish
this, we made several stacks of the deep {\em Chandra} 2 Ms X-ray data
in the GOODS-North field \citep{alexander03} for galaxies in both the
H$\alpha$ and larger UV samples, restricting the analysis to those
galaxies with spectroscopic redshifts.  The X-ray stacking analysis is
performed in a manner identical to that specified in \citet{reddy04}.
The errors on the H$\alpha$ and UV luminosities and SFRs reflect the
object-to-object dispersion in H$\alpha$/UV luminosity among the
stacked galaxies.  Examination of both the mean and median stacks of
the X-ray data suggests that there are no galaxies that dominate the
X-ray signal and bias the mean stacked results.  The results of the
stacking analysis are summarized in Table~\ref{tab:lum}, along with
conversions to SFRs listed in Table~\ref{tab:sfr}.  We will refer to
these results throughout the paper.  For the time being, we point out
that the SFRs inferred from X-ray emission are broadly consistent with
those derived from both the UV and H$\alpha$ emission once the latter
two are corrected for dust assuming the Calzetti extinction curve.
Consequently, the mean dust correction derived using the Calzetti
prescription is similar to the mean dust correction obtained by
comparing the X-ray SFRs with those derived from the observed UV and
H$\alpha$ SFRs, even when restricting the sample to those galaxies
within a specific range of SFR (Figure~\ref{fig:ha_uv}).  We will
revisit the correlation between UV slope and dust attenuation in
\S~\ref{sec:meurer}.  Based on the agreement between X-ray and
dust-corrected H$\alpha$ and UV inferences of the star formation
rates, we adopt the dust-corrected H$\alpha$ luminosity as a proxy for
the bolometric star formation rate.

\subsection{Correlation Between H$\alpha$ and $8$~$\mu$m 
Luminosity}
\label{sec:correlations}

\begin{figure}[!t]
\plotone{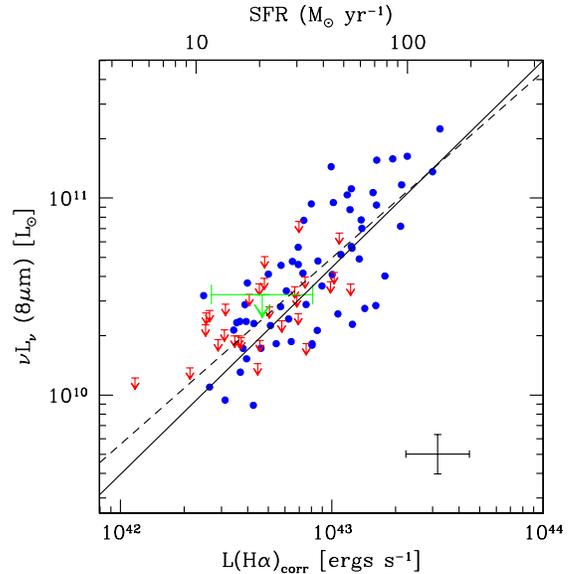}
\caption{$8$~$\mu$m versus dust-corrected H$\alpha$ luminosity for the
  90 galaxies with both $24$~$\mu$m observations and either longslit
  or narrowband measurements of the H$\alpha$ flux. The best-fit
  linear relation for the 61 detected objects ({\em circles}) is shown
  by the dashed line.  The solid line shows the fit for all galaxies,
  including the 29 undetected galaxies ({\em downward-pointing
    arrows}).  A stack of the $24$~$\mu$m data for $16$ well-isolated
  undetected galaxies yields a $3$~$\sigma$ upper limit ({\em large
    arrow}), which is measured from the background dispersion in the
  stacked image.  The width of the upper limit is determined from the
  dispersion of H$\alpha$ luminosities of these 16 galaxies.  The
  typical error on individual points is indicated in the panel.}
\label{fig:ha_l8}
\end{figure}

Figure~\ref{fig:ha_l8} shows $\mirlum$ versus dust-corrected H$\alpha$
luminosity for the 90 galaxies in our sample.\footnote{For simplicity,
  we did not consider the small fraction ($\approx 15\%$) of objects
  at $z\simeq 2.0-2.6$ targeted with NIRSPEC that did not yield
  H$\alpha$ detections.  Roughly half of the galaxies with H$\alpha$
  non-detections are also undetected at $24$~$\mu$m.  Including these
  in our analysis does little to alter the correlation between
  $\mirlum$ and dust-corrected H$\alpha$ luminosity.  Most of the
  remaining H$\alpha$ non-detected sources had inaccurate astrometry
  which may have contributed to the lack of detection.}  Focusing on
the 61 detected galaxies, we find a probability $P\simeq 0.00001$ that
the $\mirlum$ and H$\alpha$ luminosities are uncorrelated, implying a
4.4~$\sigma$ significance of the correlation.  A linear fit to the
data yields a slope consistent with unity and rms dispersion of
$0.23$~dex.  In principle, stacking the $24$~$\mu$m data for the
$24$~$\mu$m-undetected galaxies should provide additional constraints
for them.  However, unlike the Poisson-limited X-ray data, there are
additional difficulties with stacking $24$~$\mu$m data.  The first is
that the MIPS observations are typically background-limited, requiring
larger samples for a reliable stack.  The second is that owing to the
beamsize of the MIPS observations, we must ensure that we include only
well-isolated galaxies in the stack to avoid contamination from nearby
galaxies, irrespective of whether the contaminants are detected at
$24$~$\mu$m.  Unfortunately, there are not enough well-isolated
$24$~$\mu$m-undetected galaxies to obtain a reliable estimate of their
stacked flux.  Of the $29$ undetected galaxies, $16$ are at least
$3$~$\arcsec$ away from any IRAC sources.  Stacking the data for these
$16$ galaxies results in an upper limit only to their median
$8$~$\mu$m luminosity (Figure~\ref{fig:ha_l8}).  In spite of these
difficulties, we can use a survival analysis to take advantage of the
censored data (upper limits in $\mirlum$) for the undetected galaxies
to make a more robust measurement of the overall correlation between
H$\alpha$ and mid-IR luminosity.

\begin{figure*}[!t]
\plottwo{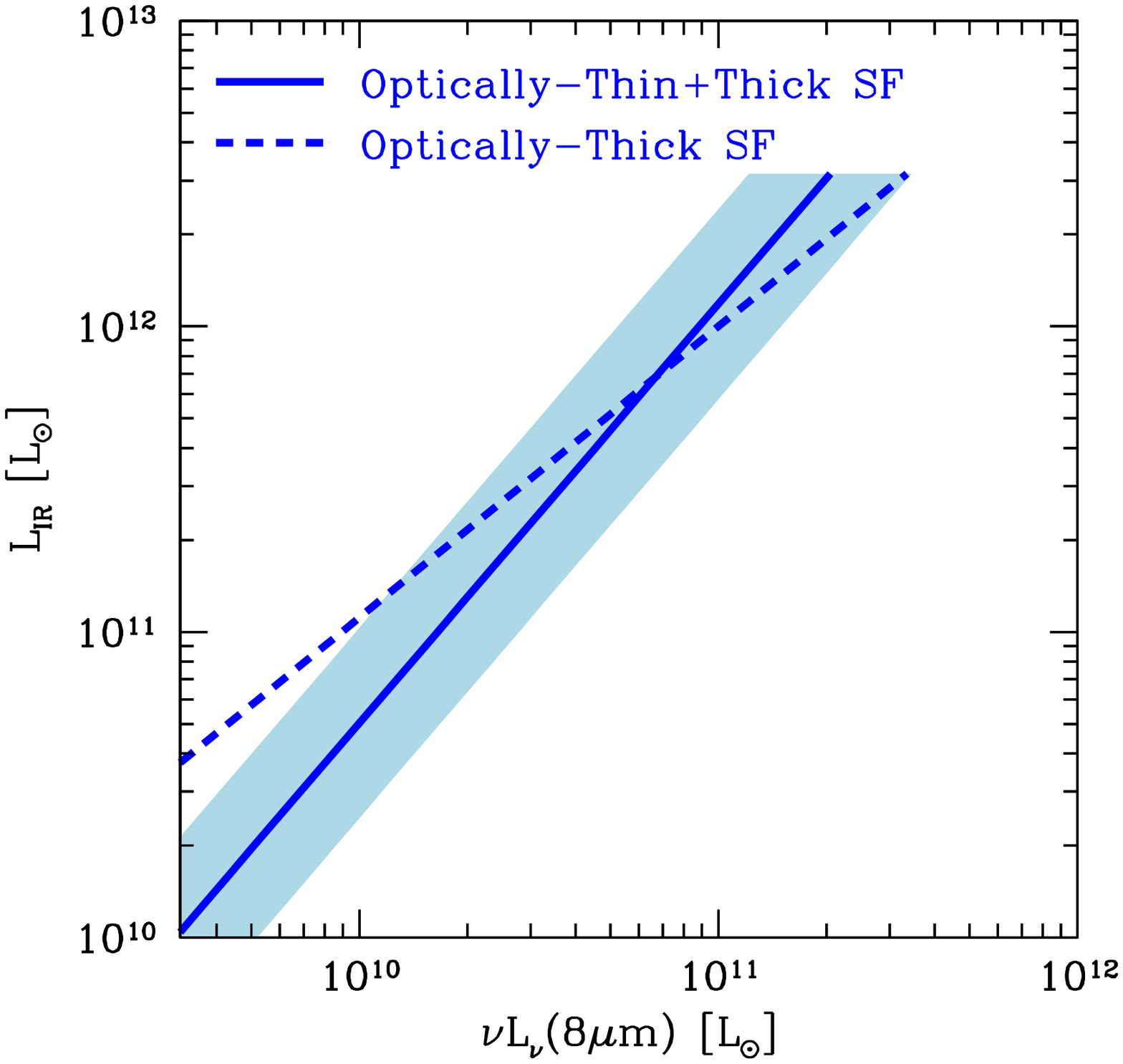}{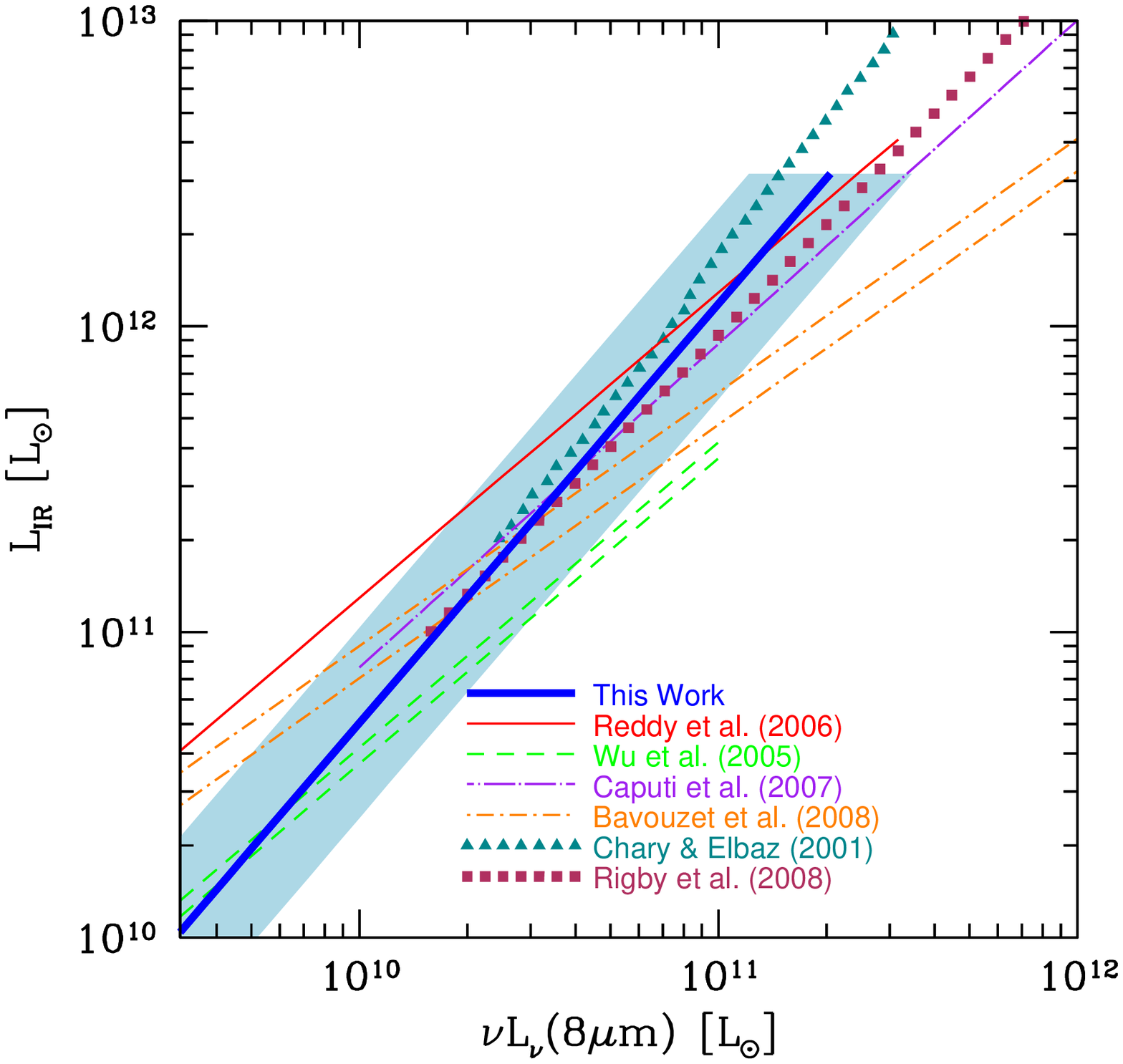}
\caption{({\em Left}): Comparison between $\lir$-$\mirlum$ relations
  in the optically thick case and in the case where the emission is
  both optically-thin and thick. $\lir$ is computed by subtracting the
  observed UV-derived SFR from the dust-corrected H$\alpha$ SFR (see
  text).  ({\em Right}): Comparison of several published calibrations
  between $\mirlum$ and $L_{\rm IR}$, with most data taken from
  \citet{caputi07}, including those of \citet{reddy06a, wu05,
    caputi07, bavouzet08, chary01, rigby08}.  The calibration and rms
  (random) dispersion derived in our work are denoted by the thick
  blue line and shaded light blue region.  We have plotted these
  relations over only the approximate luminosity range for which they
  are calibrated.}
\label{fig:l8comp}
\end{figure*}

There are several well-documented statistical techniques available for
performing linear regression in the presence of censored data,
including the Buckley-James (BJ) and expectation-maximization (EM)
algorithms, both of which are implemented in the ASURV statistical
software package \citep{isobe86}.  Survival analysis assumes that the
presence of an upper limit in a parameter (e.g., flux) is independent
of the true value of the quantity being correlated (e.g., luminosity).
This is usually satisfied in flux-limited surveys where the
fundamental parameter is luminosity.  Moreover, in our case, the
$8$~$\mu$m luminosities depend on distance in a highly non-linear way
because of the strong redshift dependence of the $k$-correction.
Finally, the upper limits are determined by the depths of the
$24$~$\mu$m images which are independent of the luminosity
distribution (i.e., we did not integrate for a longer period of time
on fainter objects).  Therefore, we assume that the $24$~$\mu$m upper
limits are roughly randomized with respect to the underlying
$8$~$\mu$m luminosity distribution.

In the EM algorithm, an inital guess is made for $\mirlum$ for the
$24$~$\mu$m-undetected galaxies based on the sample of detections and
assuming a normal distribution of residuals about the linear fit.
These guesses are then combined with the detected sample to produce a
new linear fit between $8$~$\mu$m and H$\alpha$ luminosity.  This new
fit is then used to revise the initial guesses for the luminosities of
the undetected galaxies, and the procedure is repeated until
convergence.  The BJ method works in a similar way, but makes no
assumption for the distribution of points around the regression line,
instead employing the Kaplan-Meier distribution derived from the
sample (e.g., the $24$~$\mu$m detection probability as a function of
H$\alpha$ luminosity).  Both methods were used to quantify the
relation between $\lha$ and $\mirlum$ using information for all 90
galaxies, and both methods yielded similar results.  Including both
the detections and non-detections, we find a probability $P\simeq
0.00001$ that $\mirlum$ and $\lha$ are uncorrelated, implying a
$4.4$~$\sigma$ significance of the correlation.  A linear fit to the
data yields an rms dispersion of $0.24$~dex:
\begin{eqnarray}
\log[\mirlum/L_\odot] = \nonumber \\
(1.05\pm0.11)\times\log[\lha/{\rm ergs~s^{-1}}] - (34.48\pm 4.26) \nonumber \\
{\rm for}~10^{42}\la \lha \la 4\times10^{43}~{\rm ergs~s^{-1}}.
\label{eq:l8vlh}
\end{eqnarray}
Within the errors of the fitted parameters, the linear regressions
with and without incorporating the upper limits are identical over the
range of luminosity considered here.  Note that the correlation between
$\mirlum$ and dust-corrected $\lha$ is not solely due to any
systematic change in the extinction correction with $\mirlum$ because
we still find a significant $4.0$~$\sigma$ correlation between
$\mirlum$ and observed $\lha$ (uncorrected for extinction).

Aside from the statistical analysis, it is worth examining further how
the $24$~$\mu$m-undetected galaxies affect the overall fit between
$\mirlum$ and $\lha$ based on physical grounds.  We discussed already
the limitations in stacking the $24$~$\mu$m data for these galaxies.
Therefore, we took advantage of the X-ray data for
$24$~$\mu$m-undetected galaxies in the GOODS-N field.  An X-ray stack
points to an average SFR of $\sim 20\pm 9$~M$_{\odot}$~yr$^{-1}$, very
similar to that derived from the dust-corrected H$\alpha$
(Table~\ref{tab:sfr}).  Examination of the mean attenuation factors
suggests also that while such $24$~$\mu$m-undetected galaxies are less
dusty than the detected galaxies --- as expected from the fact that
$24$~$\mu$m emission traces the dust emission in galaxies --- the
former are not completely dust-free given their relatively large SFRs
and mean attenuation of $\sim 1.3-1.5$.  Hence, the undetected
galaxies {\em cannot} have true mid-IR luminosities that fall
substantially and systematically below the correlation defined in
Eq.~\ref{eq:l8vlh}, and therefore they must be described adequately by
this correlation.

Combining the linear fit of Eq.~\ref{eq:l8vlh} with the relation
between star formation rate and H$\alpha$ luminosity
\citep{kennicutt98}, which assumes solar metallicity and does not
include the effects of stellar rotation (e.g., \citealt{leitherer08}),
we obtain the following:
\begin{eqnarray}
\log[{\rm SFR/M_\odot ~yr^{-1}}] = \nonumber \\
(0.95\pm0.10)\times\log[\mirlum/L_\odot] - (8.52\pm0.87) \nonumber \\
{\rm for}~4\times 10^9 \la \mirlum \la 2\times 10^{11}~{\rm L_\odot}.
\label{eq:l8vsfr}
\end{eqnarray}
The SFR in Eq.~\ref{eq:l8vsfr} is a {\em total} dust-corrected SFR.
In order to relate the mid-IR luminosity to total ($8-1000$~$\mu$m)
infrared luminosity, $\lir$, we cannot simply combine
Eq.~\ref{eq:l8vsfr} with the Kennicutt relation between $\lir$ and
SFR.  This is due to the fact that the latter is derived under the
optically-thick limit in which a vast majority of the bolometric
luminosity is obscured and emerges in the IR.  While this is true for
most of the galaxies in our sample, it will not be the case for
fainter ones (e.g., such as those undetected at $24$~$\mu$m) where a
large fraction of the luminosity may be unobscured (\S~\ref{sec:bol}).
The dust-corrected H$\alpha$-inferred SFR is equivalent to the sum of
the obscured (IR) and unobscured, or observed, (UV) SFRs.  Therefore,
we subtracted the observed UV SFR from the dust-corrected H$\alpha$
SFR to determine the obscured SFR and IR luminosity.  Taking the
observed UV component of the SFRs into account, we arrive at the
following relation between $\mirlum$ and $\lir$:
\begin{eqnarray}
\log[\lir/L_\odot] = \nonumber \\
(1.37\pm0.16)\times\log[\mirlum/L_\odot] - (3.01\pm1.34)\nonumber \\
{\rm for}~4\times 10^9 \la \mirlum \la 2\times 10^{11}~{\rm L_\odot}.
\label{eq:l8vir}
\end{eqnarray}
The importance of the correction for the unobscured SFR is illustrated
in Figure~\ref{fig:l8comp} where we compare Eq.~\ref{eq:l8vir} with
that obtained assuming optically-thick star formation.  Galaxies with
large $8$~$\mu$m dust luminosities have a smaller fraction of their
luminosity escaping in the UV, thus Eq.~\ref{eq:l8vir} approaches what
we would expect in the case where all of the star formation is
optically-thick.  Alternatively, galaxies with small dust luminosities
have a substantial fraction of luminosity emergent in the UV.  In this
case, the obscured SFR is lower than what we would have guessed in the
optically-thick case.  The result is that the correlation between
$\mirlum$ and $\lir$ predicts a factor of 4 lower $\lir$ than the
optically-thick case at $\mirlum \sim 3\times 10^{9}$~L$_{\odot}$.

The equations above summarize the relationship between mid-IR
luminosity, infrared luminosity, and total SFR over the ranges
specified above, as derived from the correlation between $8$~$\mu$m
and H$\alpha$ luminosity.

\subsection{Comparison with Other High-Redshift and Local Correlations}

Figure~\ref{fig:l8comp} summarizes several published determinations of
the relationship between $\mirlum$ and $\lir$, compared with our
determination at $z\sim 2$ (Eq.~\ref{eq:l8vir}).  There is a general
agreement to a factor of $\approx 2$ between the conversions over the
luminosity range where they overlap ($10^{11}\la \lir \la
10^{12}$~L$_{\odot}$).  Excepting our present determination and that
of \citet{rigby08}, all the relations shown in Figure~\ref{fig:l8comp}
were calibrated on galaxies at $z<0.7$.\footnote{\citet{bavouzet08}
  test the validity of their results for higher redshift galaxies by
  comparing with stacked $24$, $70$, and $160$~$\mu$m measurements
  (from the 3 MIPS bands) for galaxies at $z\approx 1.7$, and they
  find a mean $\mirlum$ to $\lir$ ratio similar to that of $z<0.7$
  galaxies.  Both of their fits with and without the stacked data are
  shown in Figure~\ref{fig:l8comp}.}  The calibration presented by
\citet{caputi07} is in rough agreement with our relationship within
the rms dispersion of the latter.  The \citet{wu05} relations are
calibrated on radio and H$\alpha$ data for local galaxies ($z<0.2$)
drawn from the {\em Spitzer} First Look Survey (FLS; e.g.,
\citealt{frayer06}).  Finally, the R06 calibration is derived based on
the median $\mirlum$ to $\lir$ ratio of local galaxies in the
\citet{elbaz02} sample.  Focusing on the relations calibrated with
$z\sim 2$ data, \citet{rigby08} report on the correlation between
$\mirlum$ and $\lir$ based partly on a sample of 4 lensed galaxies
with intrinsic $\lir$ between $10^{11}$ and $10^{12}$~L$_\odot$.
Their best-fit relation agrees broadly with ours over the luminosity
range where the two are calibrated.  The value of our analysis is that
it incorporates the largest sample of its kind with H$\alpha$, UV,
X-ray, and $24$~$\mu$m observations of {\em spectroscopically}
confirmed typical galaxies at $z\sim 2$.  We proceed by adopting
Eqs.~\ref{eq:l8vlh}, \ref{eq:l8vsfr}, and \ref{eq:l8vir}.

\section{RELATIONSHIP BETWEEN REST-FRAME UV SLOPE AND
DUST ATTENUATION}
\label{sec:meurer}

While the previously derived calibrations are useful, most
high-redshift galaxies are too faint or lie at too high redshift to be
detected directly at $24$~$\mu$m, thus suggesting the need for some
other proxy for dust extinction.  \citet{meurer99} demonstrated a
tight correlation between dust attenuation and rest-frame UV slope,
$\beta$, for a sample of nearby starburst galaxies.  An advantage of
using the UV slope as a proxy for dust obscuration is the ability to
quantify it for galaxies that are up to two orders of magnitude
fainter in bolometric luminosity, yet are several thousand times more
numerous, than the dustiest ultraluminous galaxies (ULIRGs) at high
redshift.  Thus, it enables us to quantify the contribution of such
galaxies to the bolometric luminosity density (e.g., \citealt{reddy08,
  reddy09}).  The Meurer relation is widely used to recover the dust
attenuation and bolometric luminosities of high-redshift galaxies,
however its validity had not been tested directly for such galaxies
until recently.  Based on a UV-selected sample of spectroscopically
confirmed $24$~$\mu$m detected galaxies and stacked X-ray analysis in
the GOODS-N field, R06 demonstrated that the local relation appears to
hold for galaxies with bolometric luminosities (sum of the IR and UV
luminosities) between $10^{11}$ and $10^{12.3}$~L$_{\odot}$ at $z\sim
2$, luminosities typical of $L^{\ast}$ galaxies at these redshifts.
Here, we re-evaluate the sensitivity of $\beta$ to dust attenuation
using the full UV-selected sample of 392 galaxies with $24$~$\mu$m
data and SED modeling.

\begin{figure*}[!t]
\plottwo{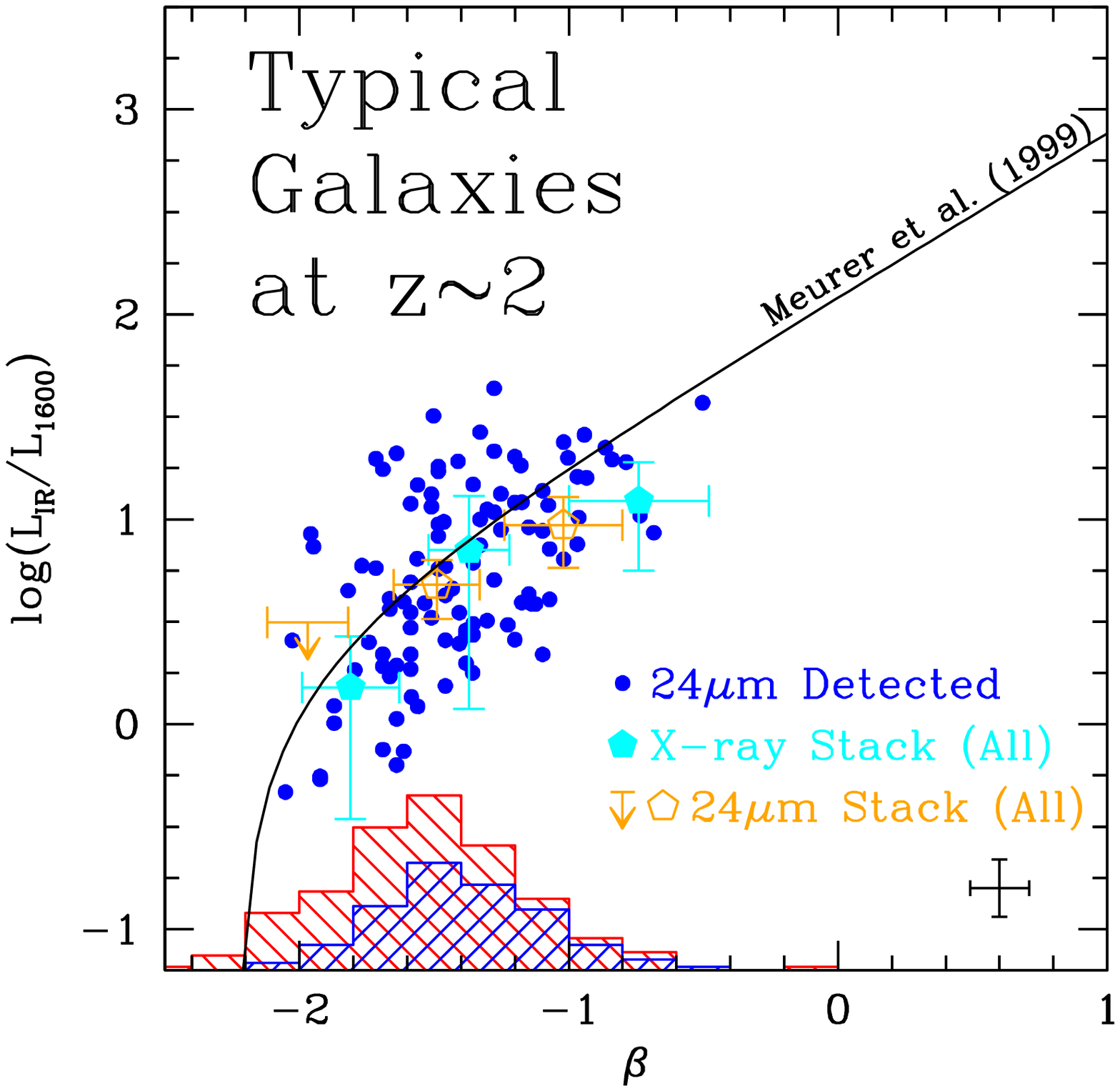}{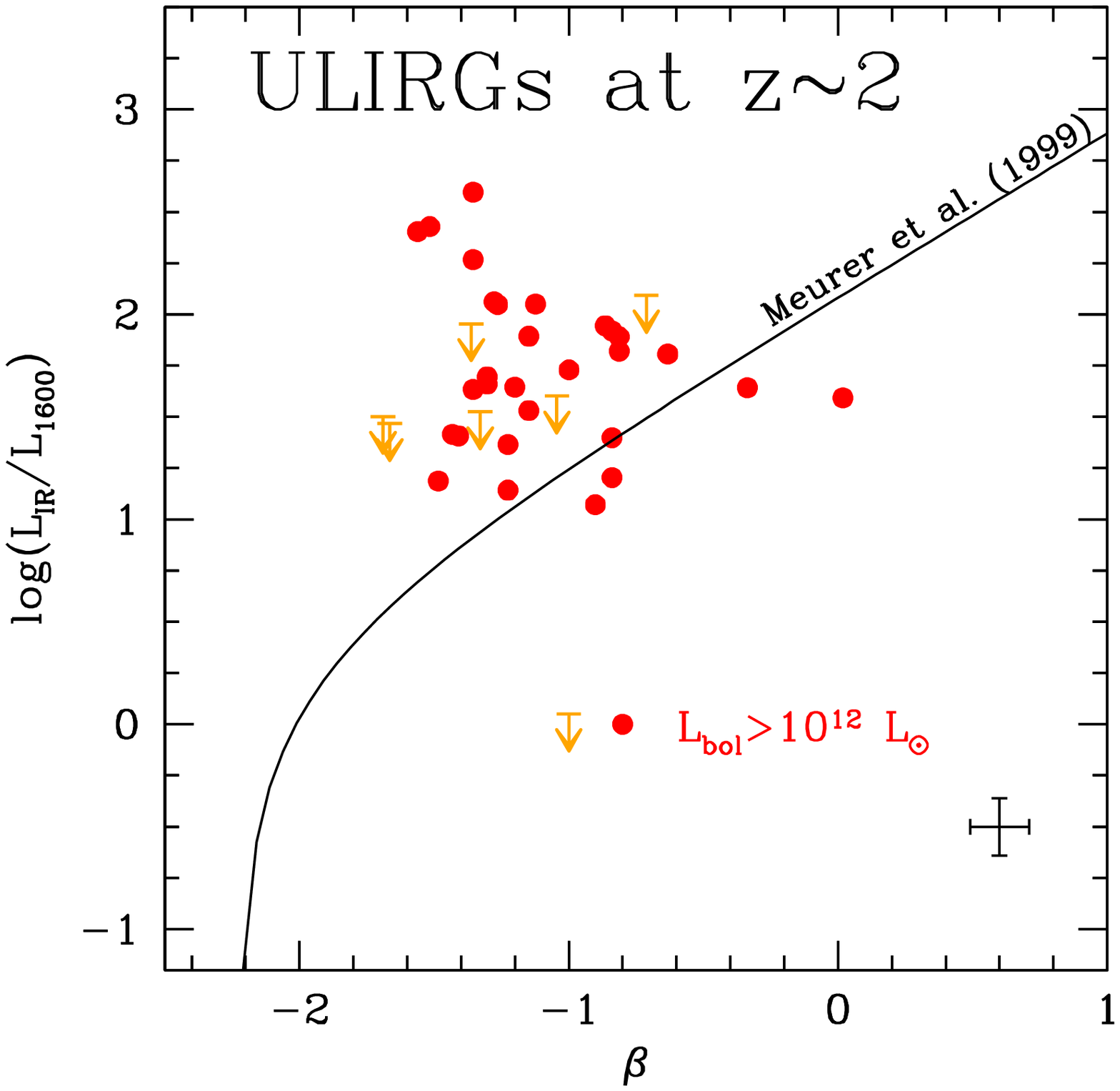}
\caption{({\em Left}): Dust attenuation, parameterized as the ratio
  between infrared and UV luminosity, versus rest-frame UV slope,
  $\beta$, for typical galaxies at $z\sim 2$ (see text).  The typical
  error in the $24$~$\mu$m detected points is indicated at the lower
  right.  The filled pentagons show our inferences from stacked X-ray
  data, where the error bars reflect the $1$~$\sigma$ dispersion in
  the relevant quantity for the stacked galaxies, which is typically
  larger than the formal uncertainty in the stacked X-ray flux.
  Similarly, the upper limit and open pentagons denote results from
  the $24$~$\mu$m stacks.  The arbitrarily normalized red and blue
  histograms show the $\beta$ distribution for galaxies undetected and
  detected, respectively, at $24$~$\mu$m.  ({\em Right}): Same as left
  panel for bolometrically luminous galaxies (ULIRGs).  Upper limits are
  shown for $24$~$\mu$m undetected galaxies.  }
\label{fig:ebmv}
\end{figure*}

\begin{figure}[hbt]
\plotone{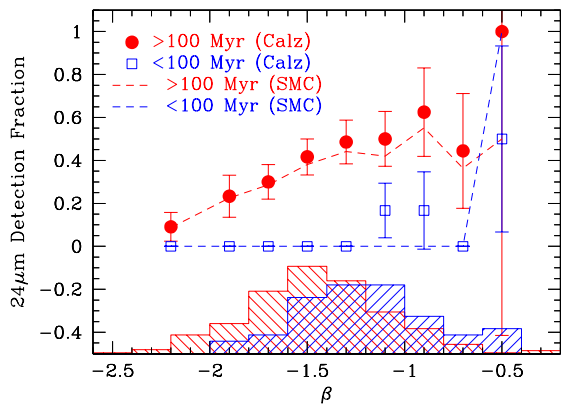}
\caption{$24$~$\mu$m detection fraction as a function of UV spectral
  slope $\beta$ for typical galaxies with ages $>100$~Myr and young
  galaxies with ages $\la 100$~Myr in the case of ages derived
  assuming a Calzetti extinction curve (open squares and solid
  circles) .  Errors assume Poisson statistics.  The dashed lines show
  the same when we assume an SMC extinction curve in deriving the
  ages.  The histograms show the arbitrarily normalized $\beta$
  distributions (in the case of Calzetti-derived ages) for the two
  subsamples, and a two-sided K-S test indicates a probability $P\la
  0.01$ that the two are drawn from the same parent distribution (see
  text).}
\label{fig:detfrac}
\end{figure}

In the subsequent discussion the dust attenuation, IRX, is
parameterized by the ratio of the infrared luminosity (computed using
Eq.~\ref{eq:l8vir}) to UV luminosity (uncorrected for dust).  The
intrinsic UV spectral slope $\beta$ is determined from the $\gmr$
color as follows.  We generated power laws in $f(\lambda)\propto
\lambda^\beta$ for $-2.5\le \beta \le 1.0$ with $\Delta\beta = 0.01$.
These are attenuated for the Ly$\alpha$ forest opacity assuming the
\citet{madau95} prescription and multiplied by the $G$ and $\rs$
transmission filters.  The UV slope is taken to be the one which gives
a $\gmr$ color closest to the observed value.  In the present
analysis, $\beta$ is simply an observed quantity and is immune to the
uncertainties associated with SED-fitting (e.g., as would be the case
with a SED-derived value of $\ebmv$).  The error in UV slope is
related directly to the error in color, and is typically $\sigma_\beta
\simeq 0.11$.\footnote{The $G$-band is affected by the Ly$\alpha$
  forest only for redshifts $z\ga 2.5$.  Therefore, statistical
  fluctuations in the forest will not affect our determination of
  $\beta$ as most of the galaxies in the samples considered here are
  at $z\la 2.5$.  For the same reason, Ly$\alpha$ contamination of the
  $G$-band flux is not a concern in our determination of $\beta$.}

\subsection{Results for Typical Star-Forming Galaxies at $z\sim 2$}

To aid our discussion, we focus first on the correlation between UV
slope and attenuation for typical star-forming galaxies at $z\sim 2$,
which we define formally as those galaxies with bolometric
luminosities, $\lbol = \lir + \luv < 10^{12}$~L$_{\odot}$, and ages
older than $100$~Myr as determined from the broadband SED fitting.
These limits are adopted to reflect the characteristics of the vast
majority of star-forming galaxies at $z\sim 2$.  Constraints on the
bolometric luminosity function imply that a typical galaxy at $z\sim
2$ will have a characteristic luminosity $L^{\ast}_{\rm bol}\sim
2\times 10^{11}$~L$_{\odot}$ \citep{reddy08}.  Further, our SED
modeling implies a median age of $360$~Myr with a dispersion of
$\approx 810$~Myr, where $\approx 13\%$ have ages younger than
$100$~Myr.  There are 311 galaxies that under this definition are
classified as ``typical,'' and 109 of these ($35\%$) are detected at
$24$~$\mu$m.  These detected galaxies exhibit UV slopes that are
correlated with attenuation ($3.8$~$\sigma$ significance) with a
formal scatter of $0.38$~dex about a linear fit
(Figure~\ref{fig:ebmv}).  The normalization of this correlation will
depend of course on the depth of the $24$~$\mu$m data, as we might
expect that the detected galaxies have dust attenuations that are
larger than those for undetected galaxies for a fixed $\beta$.  For a
fairer representation that is robust to the mid-IR observational depth
limitations, we have stacked the X-ray data for galaxies in the
GOODS-N field in three bins of $\beta$, irrespective of whether they
are detected at $24$~$\mu$m.  We convert the mean X-ray determined SFR
to an IR luminosity after taking into account the fraction of light
emerging in the UV (see discussion in \S~\ref{sec:correlations}), and
we find mean attenuation factors that exhibit the same trend with
$\beta$ as the $24$~$\mu$m-detected galaxies (Figure~\ref{fig:ebmv}).
These results are in broad agreement with those obtained by stacking
the $24$~$\mu$m data in bins of $\beta$.

More generally, the mid-IR non-detections exhibit a $\beta$
distribution that is statistically unlikely to be drawn from the same
parent distribution as the mid-IR detected galaxies.  A two-sided K-S
test indicates a probability of $0.01$ that the distribution in
$\beta$ for 24~$\mu$m detections is drawn from the same distribution
as that for the $24$~$\mu$m undetected galaxies.  While both samples
include galaxies over approximately the full range of $\beta$, those
that are undetected at $24$~$\mu$m have UV slopes that are
systematically bluer on average by $\Delta\beta\sim -0.1$ than the
detected ones.

The IRX-$\beta$ trend combined with our observations of the $\beta$
distributions for the detected versus undetected galaxies implies then
that the $24$~$\mu$m-undetected galaxies are on average less dusty
than the detected galaxies, in accord with expectations.  Further, we
find that the $24$~$\mu$m detection fraction increases steadily with
$\beta$ for typical star-forming galaxies, implying that galaxies with
redder spectral slopes are on average more infrared luminous
(Figure~\ref{fig:detfrac}).  In the next section we show that the
relationship between $\beta$ and luminosity fails for the most
luminous galaxies at $z\sim 2$.

\begin{figure*}[!t]
\plottwo{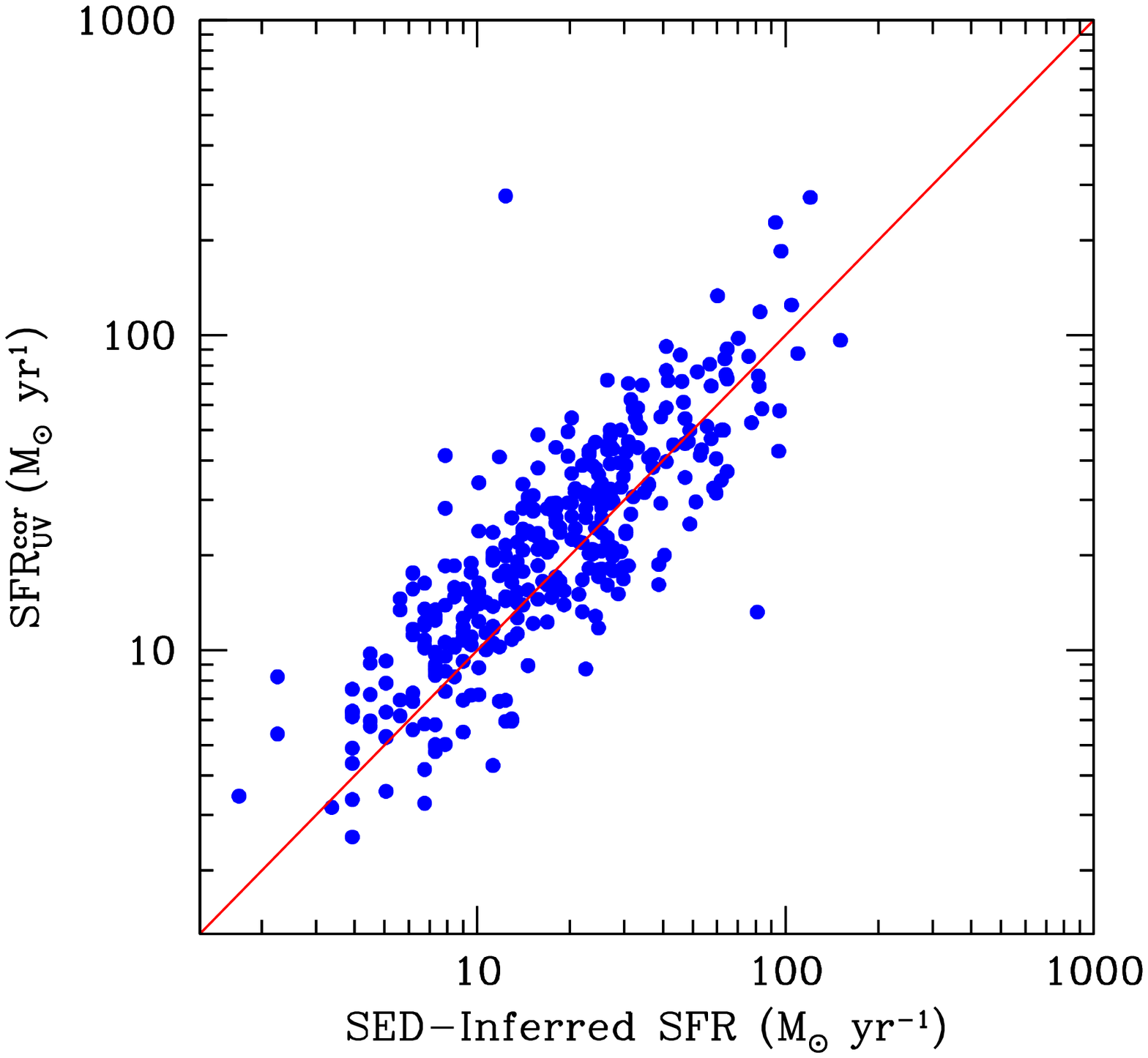}{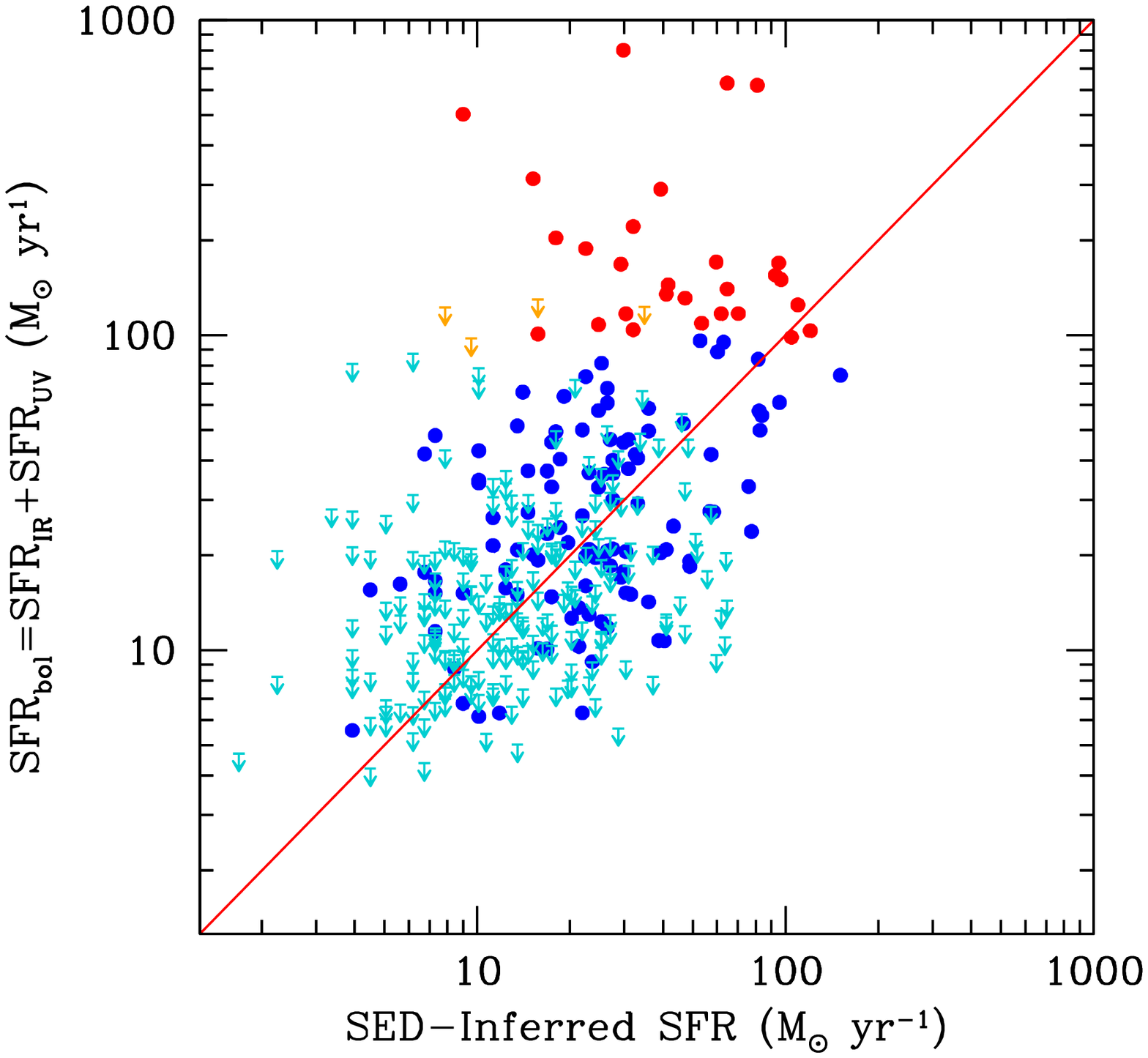}
\caption{({\em Left}): Dust-corrected UV estimate of the star
  formation rate (assuming the Meurer relation) versus the SED-derived
  star formation rates.  All SFRs assume a Chabrier IMF.  ({\em
    Right}): Bolometric SFR, defined as the sum of the IR and
  UV-derived SFRs, versus the SED-derived star formation rates.
  Orange and red points indicate galaxies with bolometric luminosities
  $\lbol\ge 10^{12}$~L$_{\odot}$.}
\label{fig:compsfr}
\end{figure*}

The correlation between attenuation and UV-slope for typical
$L^{\ast}$ galaxies at $z\sim 2$ is not necessarily surprising
considering our expectations given the correlation between mid-IR
luminosity and the H$\alpha$ luminosity corrected for dust based on
the UV slope (Figure~\ref{fig:ha_l8}).  Perhaps more importantly, the
IRX-$\beta$ relation at $z\sim 2$ is indistinguishable from that
established for local UV-starburst galaxies \citep{meurer99}, the same
relation that is almost always used to determine dust-corrected
luminosities at high redshift.\footnote{Note that \citet{meurer99} use
  a different definition of IRX than used here.  They define IRX as
  the ratio of FIR-to-UV flux density, where $\lir \approx 1.75\times
  L_{\rm FIR}$ based on the calibration of \citet{calzetti00}.  We
  have applied this correction to the Meurer relation before comparing
  to our data.}  This agreement has been noted previously for galaxies
selected using other techniques (e.g., the $BzK$-selected sample of
\citealt{daddi07a}) as well as UV-selected spectroscopic samples of
$L^{\ast}$ galaxies at $z\sim 2$ (R06; \citealt{adel00}).  Further,
the star formation rates derived by correcting the observed UV
luminosity for dust using the Meurer relation agree well with the star
formation rates inferred from the SED-fitting (left panel of
Figure~\ref{fig:compsfr}).  This result is of course not surprising
since the Calzetti extinction curve (which is essentially coincident
with the Meurer relation) is used in the SED-fitting to estimate
reddening and SFRs, both of which are driven primarily by the UV
slope.  The practical utility demonstrated here is that the UV-slope
can be used to recover the dust attenuation of typical high-redshift
star-forming galaxies to within a $1$~$\sigma$ scatter of $\approx
0.4$~dex.  This strengthens our confidence in using the Meurer
relation to recover dust attenuation from the UV SED of $L^{\ast}$
galaxies in the absence of longer wavelength data.

\subsection{Bolometrically-Luminous Galaxies}
\label{sec:ebmvbol}

Turning now to bolometrically-luminous galaxies with
$\lbol>10^{12}$~L$_{\odot}$, we find that they have systematically
larger IRX ratios than we would have inferred from the Meurer
relation.  This bias is roughly $4-5$ times larger than the random
error in the IRX ratio and is generally attributed to the fact that
significant amounts of star formation are completely obscured in the
UV (e.g., \citealt{reddy06a, papovich06, chapman05, goldader02}).
Hence, the reddening deduced from the UV SED tends to be lower than
that inferred from more direct tracers of the obscured star formation.

A pertinent question to address is whether our conversion from
$24$~$\mu$m flux to infrared luminosity is applicable for such
luminous galaxies because, as the reader will recall, our calibration
is based primarily on galaxies with $\lbol \la 10^{12}$~L$_{\odot}$.
An examination of the empirically-derived fits shown in
Figure~\ref{fig:l8comp} illustrates that, with the exception of the
Bavouzet relation, our estimate of $\lir$ for these luminous objects
is similar to what we would have predicted from the other relations.
Adopting the Bavouzet prediction would lower the IRX ratio by $\approx
0.3$~dex, an amount which is not enough to account for the mean
$1$~dex offset of $>10^{12}$~L$_{\odot}$ galaxies from the Meurer
relation.  Further, as we show in \S~\ref{sec:bol}, the relationship
between bolometric luminosity and dust attenuation implies that
ultraluminous infrared galaxies (ULIRGs) with $\lbol \ga
10^{12}$~L$_{\odot}$ on average will have fainter observed UV
luminosities, and thus IRX ratios that are boosted, relative to LIRGs.
Consequently, the bias of galaxies with $\lbol > 10^{12}$~L$_{\odot}$
to have UV slopes that systematically underestimate their attenuation
is likely to be a physical effect, as opposed to one driven by
luminosity-dependent biases in the calibrations between mid-IR and
total IR luminosity.

\begin{figure}[hbt]
\plotone{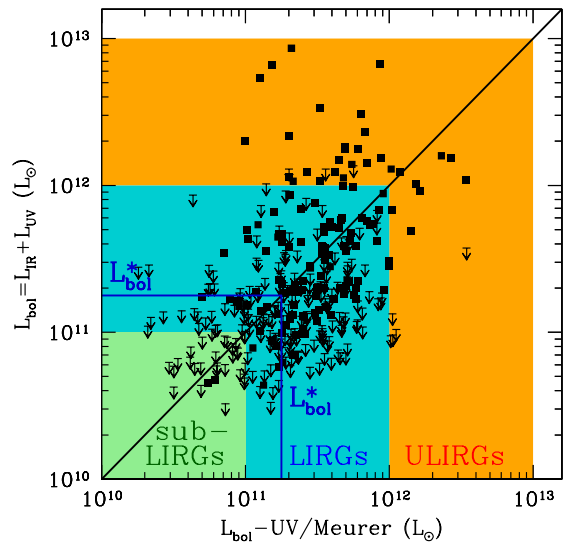}
\caption{Comparison between the direct tracer of bolometric luminosity
($\lbol = \lir + \luv$) and that obtained by combining $\luv$ with the
Meurer prediction of the dust attenuation.}
\label{fig:bolcomp}
\end{figure}

Deviation of the these ultraluminous galaxies from the Meurer relation
can be seen more clearly in Figure~\ref{fig:bolcomp}, where we compare
the direct tracer of bolometric luminosity (sum of infrared and UV
luminosities) with that obtained when we combine the UV luminosity
with the Meurer prediction for the dust attenuation.  Over the
luminosity range probed by the data, the Meurer relation successfully
predicts the bolometric luminosity for LIRGs, but significantly
underpredicts the luminosity of ULIRGs.  This naturally translates to
a disagreement between the SED-derived SFRs and those derived from
summing the IR and UV-based SFRs for ultraluminous galaxies at these
redshifts, since the SED-derived SFRs assume the Calzetti extinction
curve (right panel of Figure~\ref{fig:compsfr}).  The critical point
is that the Meurer relation can be used reliably up to $\lbol \simeq
10^{12}$~L$_{\odot}$ at $z\sim 2$, and thus is valid for typical
star-forming galaxies at these redshifts.

\begin{figure*}[!t]
\plottwo{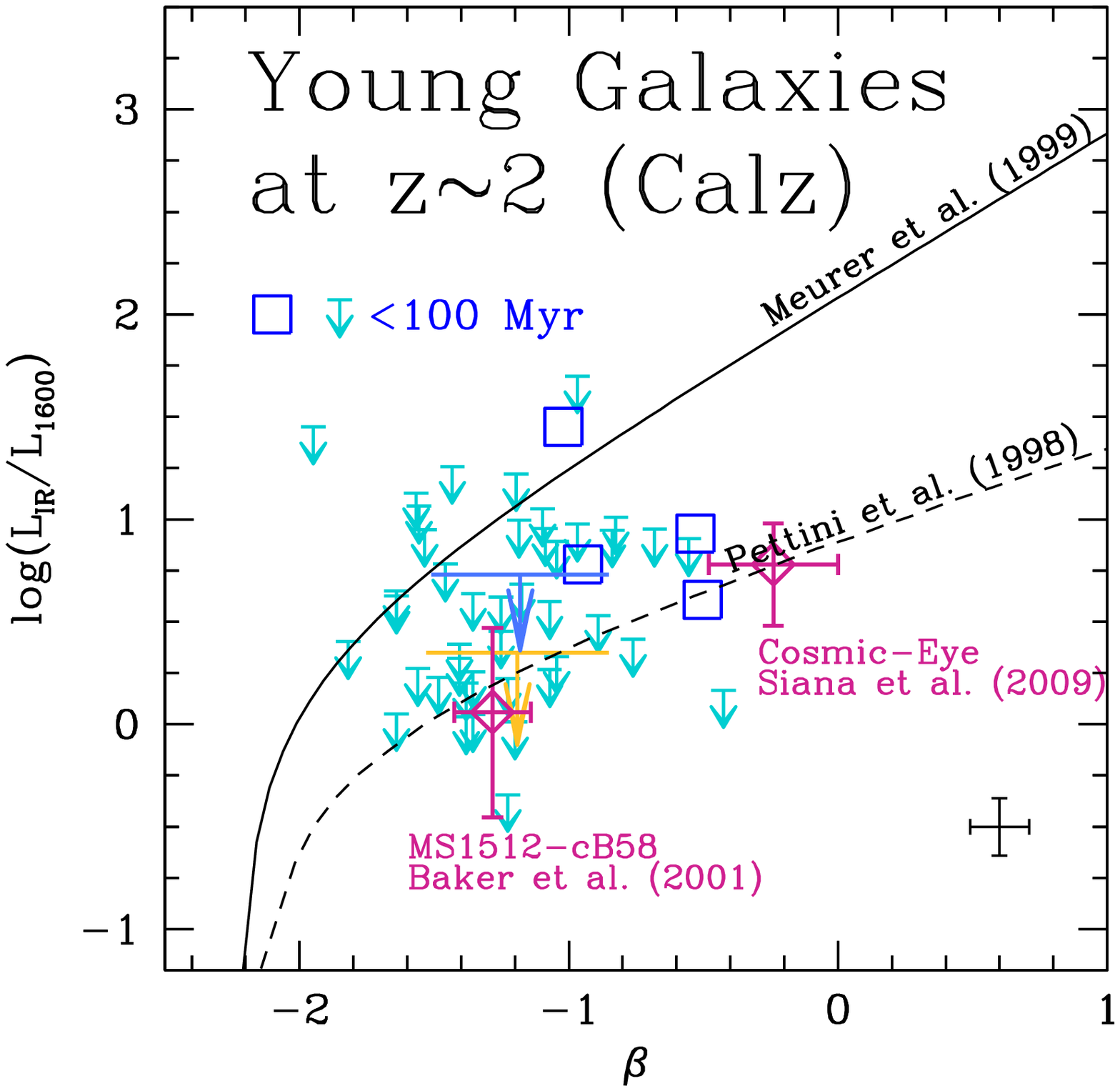}{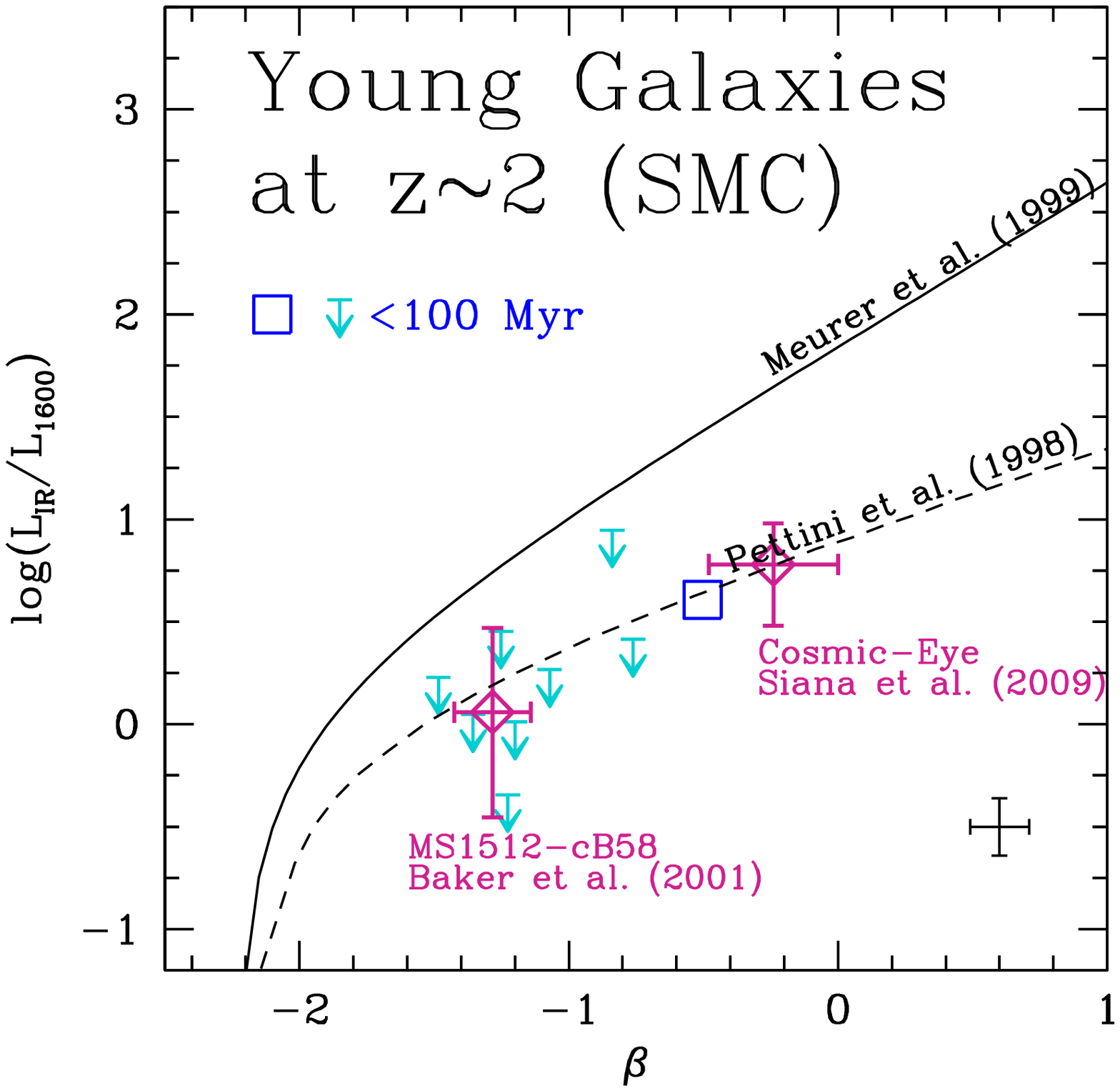}
\caption{Same as Figure~\ref{fig:ebmv} for galaxies younger than
  $100$~Myr, where ages are determined from stellar population
  modeling assuming the Calzetti ({\em left panel}) and SMC ({\em
    right panel}) extinction curves.  The open squares and small
  downward pointing arrows denote galaxies detected and undetected,
  respectively, at $24$~$\mu$m.  The large light blue and yellow
  downward pointing arrows denote the $5$~$\sigma$ limits from stacks
  of the $24$~$\mu$m and X-ray data, respectively, for the $<100$~Myr
  galaxies.  Also shown are direct measurements based on CO and {\em
    Spitzer} IRS observations of the two lensed LBGs, MS1512$-$cB58
  and the Cosmic Eye, and the form of the SMC extinction curve
  presented in \citet{pettini98}.  There are fewer points in the right
  panel since ages determined with the SMC curve are generally older
  than those derived with the Calzetti curve.  Thus, the fraction of
  galaxies considered ``young'' ($<100$~Myr) is significantly smaller
  for the SMC case relative to the Calzetti case (see text).}
\label{fig:ebmvyoung}
\end{figure*}

\subsection{Young Galaxies with Inferred Star-Formation Ages of $\la 100$~Myr}
\label{sec:ebmvyoung}

We now turn to the 49 galaxies with ages $\la 100$~Myr; this young
sample constitutes $\approx 13\%$ of the sample of 392 galaxies.  Only
4 of the 49 galaxies are detected directly at $24$~$\mu$m
(Figure~\ref{fig:ebmvyoung}).  About three fourths of the young
galaxies undetected at $24$~$\mu$m have limits in IRX that imply that
they lie below the Meurer relation.  Stacking the $24$~$\mu$m data
results in a conservative $5$~$\sigma$ limit in IRX that suggests that
these young galaxies are in general less attenuated than their older
counterparts at a fixed value of $\beta$.  This result is further
supported by a stack of the X-ray data for the 31 young galaxies in
the GOODS-N field that places a $5$~$\sigma$ upper limit in IRX that
is still 0.5~dex lower than the Meurer expectation.

These observations suggest that an SMC-like (as opposed to a Calzetti)
extinction curve may be more appropriate in describing young galaxies
in our sample.  Because the ages are derived from the stellar
population modeling (\S~\ref{sec:sedfit}) and are therefore degenerate
with respect to the assumed extinction curve (Calzetti), it seems
prudent to determine how the inferred ages are perturbed under the
assumption of an SMC extinction curve.  Adopting an SMC curve will
generally yield older ages relative to the Calzetti assumption because
in the SMC case a smaller fraction of the optical minus near-infrared
color is attributed to dust and a larger fraction is attributed to an
older stellar population (e.g., \citealt{shapley01}).

\begin{figure}[h]
\plotone{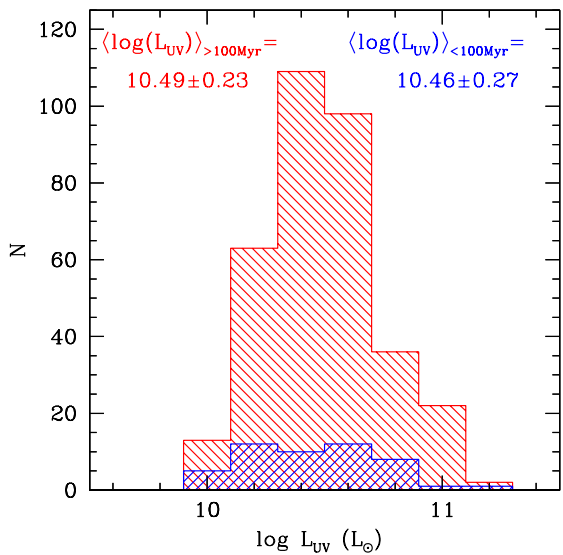}
\caption{UV luminosity distribution of galaxies with ages less than
and older than 100 Myr.}
\label{fig:luvdist}
\end{figure}

The results obtained by adopting an SMC curve are summarized in
Figure~\ref{fig:ebmvyoung}.  As expected, the fraction of galaxies
considered ``young'' ($\la 100$~Myr) is a factor of 5 times lower
($9/392$) than that obtained with a Calzetti curve.  More relevant to
our discussion here is that even when considering an SMC curve, we
note a significant departure of the youngest galaxies from the locus
of points that are characteristic of older galaxies in the IRX-$\beta$
plane that lie on the Meurer relation.  At face value, these arguments
may seem a bit circular given the degeneracy between the assumed
extinction law and inferred ages.  However, both IRX and $\beta$ are
determined {\em independently} of the stellar population modeling and
Figure~\ref{fig:ebmvyoung} demonstrates clearly that irrespective of
the choice of extinction curve, the IRX-$\beta$ properties of the
youngest galaxies deviate from those of older, more typical ones at
$z\sim 2$ (c.f., Figure~\ref{fig:ebmv}).

\subsubsection{$24$~$\mu$m Detection Fraction and $\beta$ Distribution}

The differences between the young and old populations are demonstrated
more clearly in Figure~\ref{fig:detfrac} where we show the $24$~$\mu$m
detection fraction for the young galaxies versus older more typical
star-forming galaxies.  While the detection fraction for typical
star-forming galaxies increases steadily with $\beta$, that same
fraction for the young galaxies stays essentially flat (with no
detected objects).  The only young galaxies detected directly at
$24$~$\mu$m are those with the reddest spectral slopes.  Because the
redshift and UV luminosity distributions of galaxies in the young and
typical subsamples are not substantially different
(Figure~\ref{fig:luvdist}), the trend between detection frequency and
UV slope implies that the youngest galaxies are less infrared-luminous
and less dusty on average than typical galaxies at a fixed $\beta$.
Note further that the distribution of spectral slopes for the older
and younger subsamples are unlikely to have been drawn from the same
parent distribution.  A K-S test implies a probability $P\simeq 0.005$
that the $\beta$ distribution of young galaxies is drawn from the same
parent distribution as that of the older galaxies.  Thus, our analysis
implies that the youngest galaxies are not only less infrared-luminous
and less dusty on average than older galaxies with the same range in
$\beta$, but they also tend to have redder spectral slopes than older
galaxies.  The difference in mean $\beta$ between galaxies with ages
$\le 100$~Myr and $>100$~Myr is $\langle \beta\rangle_{\rm \le
  100~Myr} - \langle \beta\rangle_{\rm > 100~Myr} = 0.26$.  The
apparently contradictory results that the young galaxies have both
redder spectral slopes {\em and} are on average less dusty than
typical galaxies may be understood if the reddening curve of the young
galaxies is ``steeper'' (i.e., less dust is required to redden the
continuum) in the UV than the typically assumed Meurer/Calzetti
relations, as alluded to above.  To rule out any biases that may
underlie the offset of these galaxies from the Meurer relation, we
must scrutinize the validity of the measures of attenuation for such
young galaxies, as we discuss next.

\subsubsection{Measures of $\lir$ in Young Galaxies}

There are several independent reasons to suspect that the $\lir$ we
compute for these young galaxies are not biased by the way in which we
are measuring them.  We consider first the possibility that galaxies
with ages $\la 100$~Myr have systematically lower $\mirlum/\lir$
ratios than their older counterparts.  Specifically, it is thought
that much of the carbon dust in the ISM is ejected from AGB stars on
timescales of $\simeq 200$~Myr, whereas the larger dust grains
responsible for the bulk of the IR emission are produced on the much
shorter timescale of Type II supernovae.  Local observations appear to
support this scenario in the sense that PAH metallicity (Z$_{\rm
  PAH}$) is observed to be significantly lower relative to the IR dust
grain metallicity (Z$_{\rm dust}$) for metal poor galaxies, but that
the ratio of the metallicities increases rapidly until an equilibrium
is reached for gas phase metallicities (Z$_{\rm gas}$) in excess of
$\approx 0.3$~Z$_{\odot}$, owing to increased carbon ejection by AGBs
\citep{galliano08}.  This delayed injection of carbon grains results
in a depressed ratio of mid-IR to total-IR flux for galaxies younger
than $200$~Myr.  However, direct measurements of the mid-IR spectral
features from {\em Spitzer} IRS spectra and long wavelength ($\ga
70$~$\mu$m) constraints on the infrared luminosities of two young
($\la 300$~Myr) lensed LBGs at $z\sim 3$ (MS1512$-$cB58 and the
``Cosmic Eye'') imply $\mirlum/\lir$ ratios that are similar to those
of local star-forming galaxies and high-redshift submillimeter
galaxies \citep{siana08, siana09}, and older LBGs at $z\sim 2-3$
\citep{reddy06b}.  In addition, an independent deduction based on CO
observations of MS1512$-$cB58 \citep{baker01} points to an $\lir$ that
is not significantly different than what we would have predicted from
our calibration between $\mirlum$ and $\lir$.  Further, if we restrict
the local correlation between Z$_{\rm PAH}$, Z$_{\rm dust}$, and
Z$_{\rm gas}$ to the same dynamic range in gas-phase metallicity as
observed among LBGs at $z\sim 2$, $0.3\la$ Z$_{\rm gas} \la$
Z$_{\odot}$ \citep{erb06a}, we would find a PAH to dust grain
metallicity ratio that is roughly constant.  The limited dynamic range
in gas-phase metallicity probed by the current sample, as well as the
direct measurements of the PAH to IR ratio in lensed LBGs at $z\sim
3$, suggests that our conversion between $\mirlum$ and $\lir$
(calibrated primarily on galaxies older than $100$~Myr) should be a
reasonable approximation for the younger galaxies as well.  In this
case, the mid-IR emission observed in the young galaxies may be due to
some combination of silicate and amorphous carbon dust produced in
Type II supernovae \citep{todini01}.

Finally, the X-ray data provide an independent confirmation of our
results.  The X-ray emission in starburst galaxies arises primarily
from shock-heated outflowing gas \citep{strickland04, grimes05,
  grimes06} and high-mass X-ray binaries \citep{ghosh01}, both of
which should be sensitive to star formation on relatively short
timescales ($\simeq 70$~Myr) that are comparable to the dynamical
timescale noted above.  Consequently, if the true infrared
luminosities of the young galaxies were larger than what we infer from
$\mirlum$ (and if $\lir$ was consistent with the Meurer prediction),
then the X-ray stack of the 49 young galaxies should have yielded a
significant detection.  Yet, we are able to place a firm $5$~$\sigma$
upper limit on the the X-ray-inferred IRX of $\log(\lir/\luv) < 0.35$.

In summary, galaxies with ages $\la 100$~Myr appear to have lower
attenuation on average than their UV slopes would imply from the
Meurer relation (note that the Meurer and Calzetti relations track
each other closely for $\beta>-2.0$).  This inference is based on the
limited dynamic range in metallicity probed by UV-bright ($\rs \la
25.5$) LBGs, direct measurements of $\mirlum$ and $\lir$ in at least a
couple of young lensed LBGs, stacked X-ray measurements, and the
$24$~$\mu$m detection fraction as a function of UV slope.  Similar
conclusions are reached by R06 and \citet{siana08, siana09}.  We have
shown here that this result appears to apply generically to most young
$\la 100$~Myr galaxies at $z\sim 2$.  The small fraction of young
galaxies that have red $\beta$ and are detected at $24$~$\mu$m points
to a scatter in IRX for young galaxies that may be larger than for
their older counterparts.\footnote{For example, applying the Meurer
  relation to one young LBG at $z=2.83$ (``Westphal-MM8'') results in
  a bolometric luminosity similar to that inferred from its detection
  at $850$~$\mu$m \citep{chapman09}.}  Direct detection of the dust
emission of a statistical sample of young high-redshift galaxies is
required to accurately constrain their scatter in attenuation.

We conclude by noting the following.  First, only 10 of the 90
galaxies used to constrain the relation between $\mirlum$ and $\lha$
have (Calzetti-derived) ages $<100$~Myr, and redetermining the
relation excluding these 10 sources does little to affect the overall
fit between $\mirlum$ and $\lha$.  In other words, our assumption of
the Calzetti relation in dust correcting the $H\alpha$ luminosities
for these sources minimally affects our conversion between $\mirlum$
and $\lir$.  Second, when computing bolometric SFRs, the
overestimation of attenuation by applying the Meurer/Calzetti
relations may be partly compensated for by the fact that the UV
luminosities in young galaxies will {\em underpredict} the star
formation rate.  The latter effect is due to the fact that for a
constant star formation history the ratio of O and B stars
contributing to the UV continuum will stabilize only after the main
sequence lifetime of B stars of $\simeq 100$~Myr.  This underscores
the need to correct for both a different attenuation and a different
conversion between UV luminosity and SFR when inferring the total star
formation rates of young galaxies (ages $\la 100$~Myr) at high
redshift.

\begin{figure*}[!t]
\plottwo{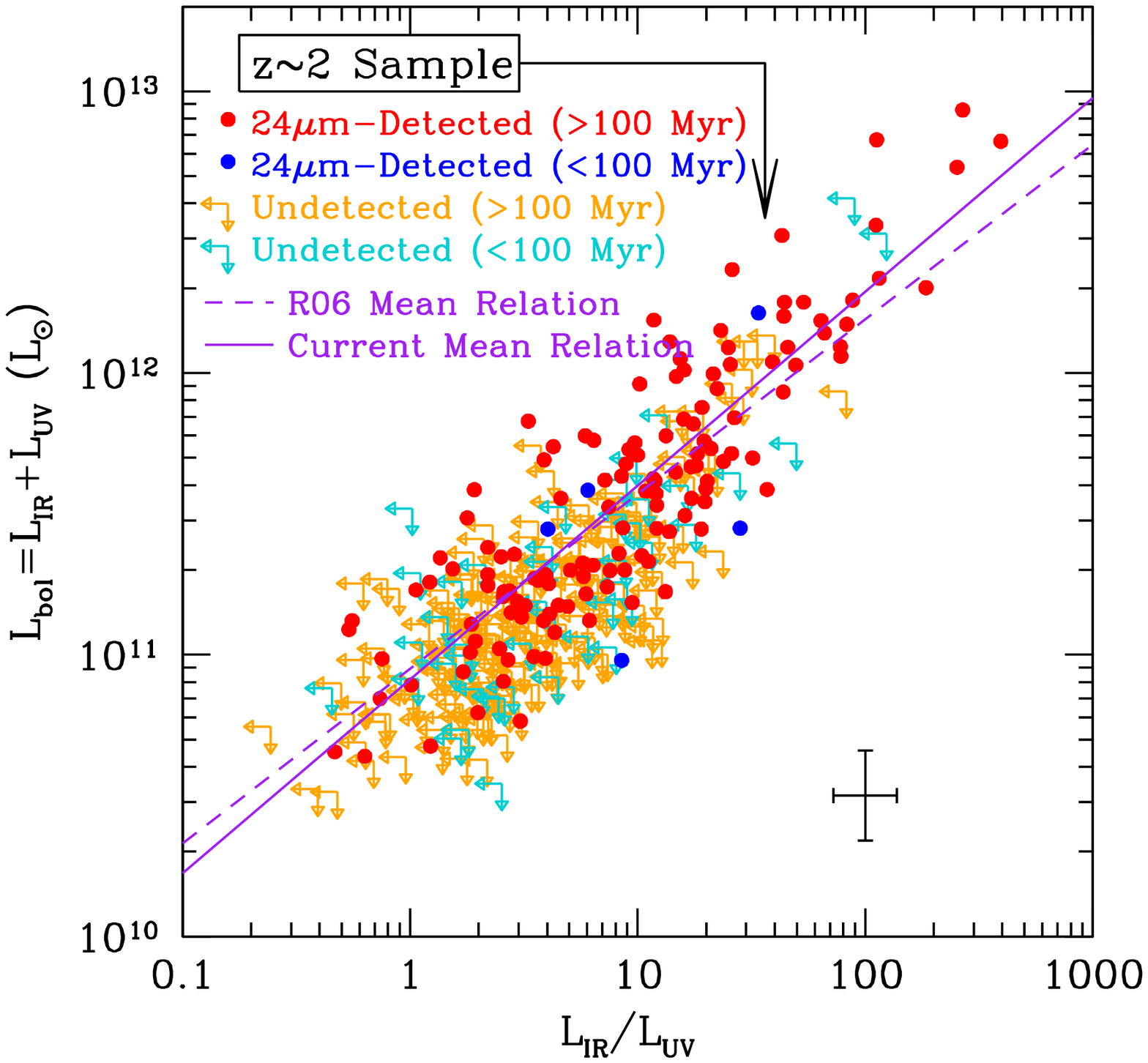}{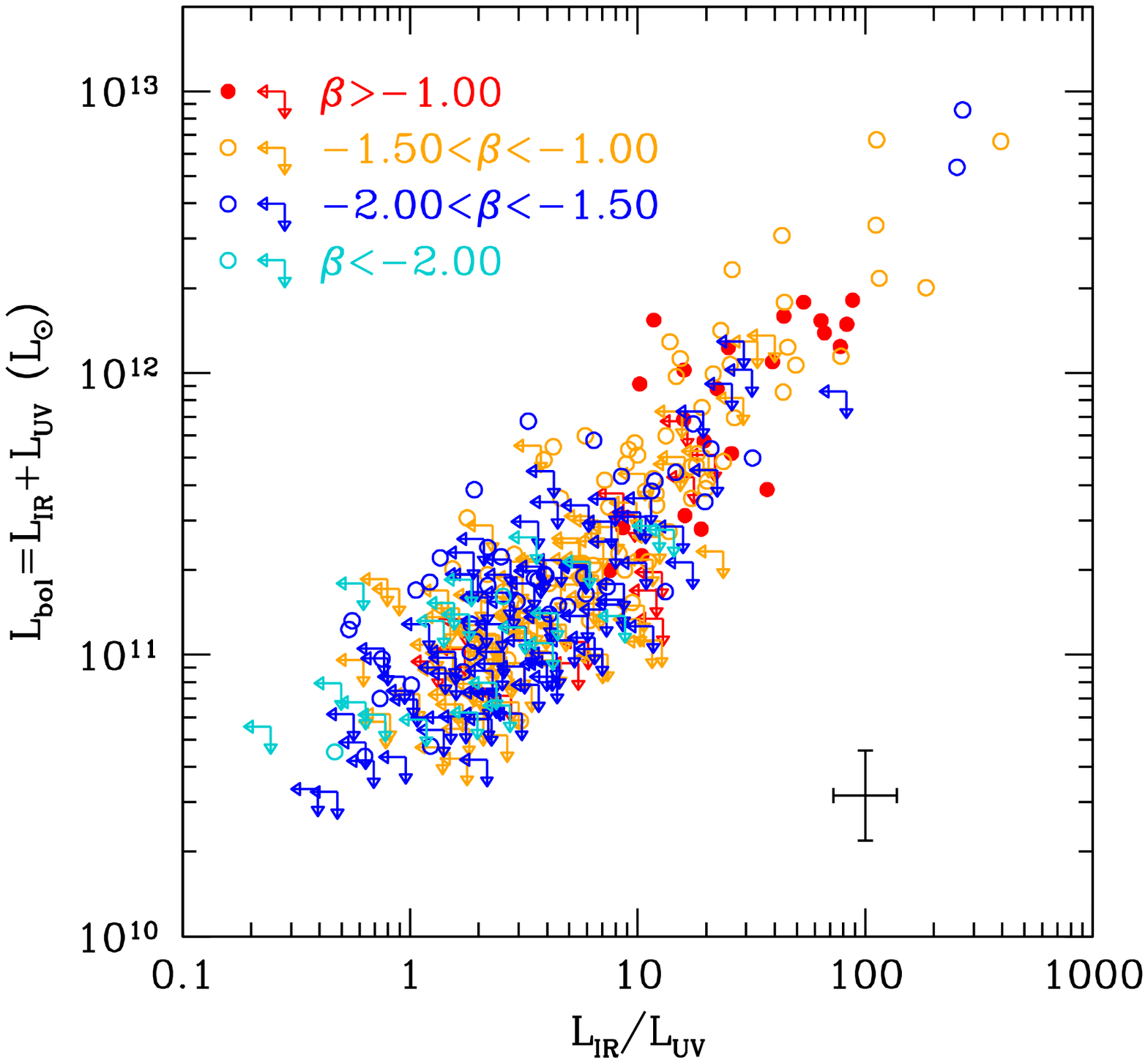}
\caption{({\em Left}): Bolometric luminosity versus dust attenuation
  for the sample of 392 galaxies with spectroscopic redshifts between
  $1.5<z<2.6$.  ({\em Right}): Same as {\em left} panel but color
  coded by UV slope $\beta$ and restricted to galaxies with ages
  $>100$~Myr.}
\label{fig:bolfig}
\end{figure*}

\subsection{Summary}

We have used the relationship between $\mirlum$, SFR, and $\lir$ to
examine the variation of UV slope with dust attenuation for typical
galaxies, bolometrically-luminous ones, and the youngest galaxies at
$z\sim 2$.  We find a significant correlation between UV slope and
attenuation for the vast majority of typical star-forming galaxies at
$z\sim 2$.  In this case, the UV slope can be used to recover dust
attenuation to within a scatter of 0.4~dex.  Those galaxies with
bolometric luminosities in excess of $10^{12}$~L$_{\odot}$ have dust
obscurations that exceed by almost a decade those predicted from the
UV slope, owing to the larger fraction of obscured luminosity in these
galaxies and, as we show in \S~\ref{sec:bol}, a decrease in observed
UV luminosity relative to galaxies with moderate bolometric
luminosities.

Finally, for the $\la 13\%$ of our sample (exact fraction depends on
the extinction law used to model the galaxies' photometry; see
discussion above) that consists of young galaxies with ages $\la
100$~Myr, the local correlation overpredicts dust attenuations at a
given $\beta$.  This implies that such young galaxies may on average
follow an extinction curve that deviates from the usually assumed
Meurer/Calzetti.  These results are similar to what we found
previously based on a smaller sample of galaxies in the GOODS-North
field (R06).  Here, we have expanded these initial results by
exploring in more detail some of the systematics that may give rise to
such an offset.  We find that the undetected young galaxies and the
two lensed LBGs from the literature exhibit $\beta$ and IRX that are
consistent with an SMC dust extinction curve (e.g.,
\citealt{pettini98}).  This behavior may be due to a difference in
covering fraction of dust, where the younger galaxies have larger
covering fractions consistent with a foreground screen of dust as
described by an SMC-like extinction curve.  Alternatively, we cannot
rule out the possibility that the dissimilar extinction curve for
young galaxies may arise from differences in the dust grain size
distribution.  This may be the case if the dust giving rise to the
mid-IR emission in young galaxies is produced primarily in Type II SNe
and contrasts in size and composition from dust produced in lower mass
stars (e.g., \citealt{maiolino04, todini01}; see also discussion in
\citealt{siana09}).

\section{Relationship between Bolometric Luminosity and Dust Attenuation}
\label{sec:bol}

Employing our measures of the dust obscuration and bolometric
luminosities for typical star-forming galaxies at $z\sim2$, we proceed
to examine the relationship between these two quantities, its
implication for the correlation between UV and bolometric luminosity,
and its redshift evolution.

\begin{figure*}[hbt]
\plotone{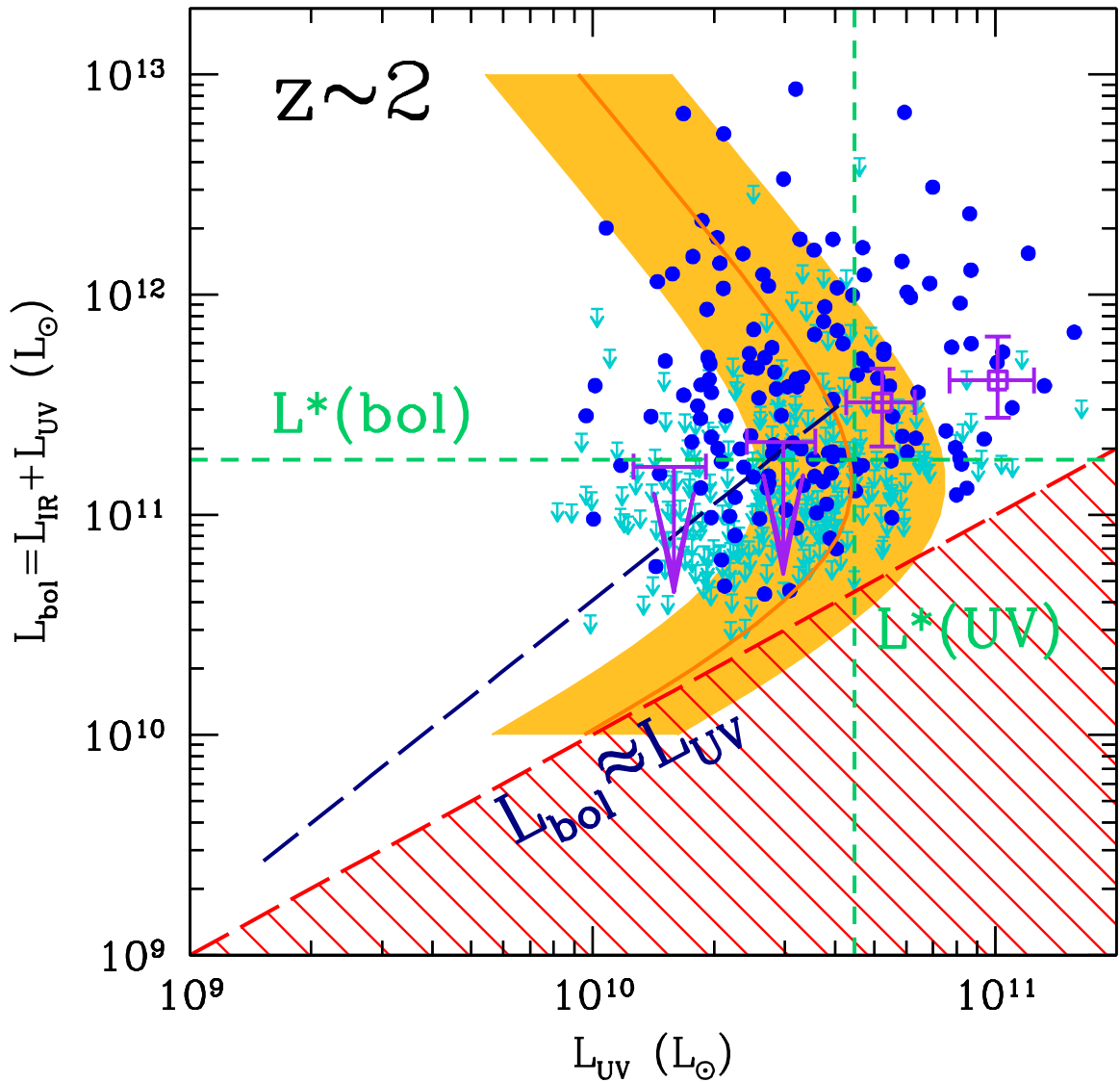}
\caption{Bolometric luminosity ($\lbol$) versus observed UV luminosity
  ($\luv$).  Small points and arrows indicate galaxies detected and
  undetected, respectively, at $24$~$\mu$m. The $24$~$\mu$m stacked
  results are shown by the open squares and large arrows ({\em
    purple}).  The shaded region denotes $\pm 1$~$\sigma$ about the
  mean relation implied by the correlation between $\lbol$ and dust
  attenuation (Eq.~\ref{eq:boleq}).  The hashed region indicates the
  area excluded by the fact that $\lbol$ must be greater than $\luv$.
  The dashed horizontal and vertical lines delineate the values of
  $L^{\ast}_{\rm UV}$ and $L^{\ast}_{\rm bol}$ at $z\sim 2$
  \citep{reddy09, reddy08}, and the thicked dashed line shows the
  extrapolation of the relation to UV-faint galaxies (see text).}
\label{fig:bolvuv}
\end{figure*}
 
\subsection{Functional Form and $\beta$ Dependence}

Based on the sample of 392 galaxies with MIPS observations, we find a
strong correlation between $\lbol$ and dust attenuation with a formal
scatter of 0.23 dex about a linear fit (Figure~\ref{fig:bolfig}).  The
fit to the present data is essentially identical to the linear fit to
a smaller sample in the GOODS-N field (R06), despite the differences
in the conversion between $\mirlum$ and $\lir$ used in the current
versus previous (R06) study.  This similarity (and the small scatter)
is due to the fact that $\lbol = \lir + \luv$ is highly correlated
with dust attenuation, $\lir/\luv$, according to our definitions.
Nonetheless, we are confident of our calibration between bolometric
luminosity and dust attenuation given that $\lbol$ estimated from the
sum of the IR and UV luminosities is consistent with $\lbol$ (or SFR)
estimated from X-ray data and from the dust-corrected H$\alpha$ and UV
luminosities (\S~\ref{sec:stack}).  Further, R06 showed
that the relationship between $\lbol$ and $\lir/\luv$ for UV-selected
samples remains valid for star-forming galaxies selected on their
rest-frame optical colors and/or submillimeter emission.  More
recently, \citet{huang09} find that IRAC-selected ULIRGs at $z\sim
1.9$ lie on the same relation as defined by the typically less
luminous UV-selected galaxies (LIRGs).  These results imply that for
the observationally-accessible area of the $\lbol$-$\lir/\luv$ plane,
the relationship defined by UV-selected galaxies is not substantially
different or biased with respect to that defined by galaxies selected
by other means (e.g., optical and IR selections).  Formally, our
best-fit relation between $\lbol$ and dust obscuration is
\begin{eqnarray}
\log[\lbol/L_{\odot}] = \nonumber \\
(0.69\pm0.03)\log[\lir/\luv] + (10.91\pm0.04) \nonumber \\
{\rm for}~\luv \ga 10^{10}~{\rm L_\odot}.
\label{eq:boleq}
\end{eqnarray}
Similar relations between bolometric luminosity, or SFR, and dust
attenuation have been found at low and high redshifts \citep{wang96,
  adel00, reddy06a, buat07, burgarella08, buat09}.

The variation of spectral slope with attenuation and $\lbol$ can be
seen in the right panel of Figure~\ref{fig:bolfig} where points are
color coded by $\beta$.  Galaxies with red $\beta>-1.00$ are on
average more attenuated and more bolometrically luminous than galaxies
with bluer $\beta$.  Galaxies with $-1.50<\beta \le -1.00$ span a
relatively larger range of attenuation and $\lbol$, reflecting the
fact that many bolometrically luminous galaxies have a similar range
in $\beta$ (but much larger attenuation) than less luminous galaxies
(\S~\ref{sec:ebmvbol}).  Galaxies with the bluest $\beta$ span a
narrower range in attenuation and $\lbol$, both being on average lower
for these galaxies.

\begin{figure}[!t]
\plotone{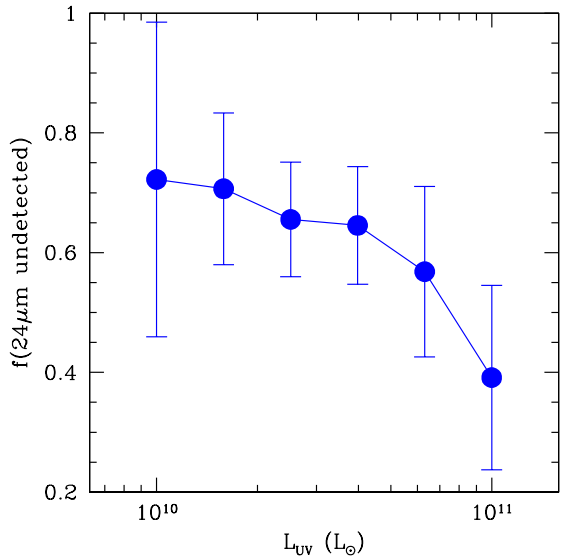}
\caption{$24$~$\mu$m non-detected fraction as a function of unobscured
  UV luminosity ($\luv$).  Error bars assume Poisson statistics.}
\label{fig:luvundet}
\end{figure}

\subsection{Variation of Observed UV Luminosity with Bolometric Luminosity}
\label{sec:bolvuv}

\subsubsection{Bolometric Luminosities of UV-Faint Galaxies}
\label{sec:bolfaint}

For further insight, we have recast the correlation between bolometric
luminosity and dust attenuation (Eq.~\ref{eq:boleq}) in terms of the
observed UV luminosity ($\luv$),
\begin{eqnarray}
\log[\luv/L_{\odot}] = \nonumber \\
\log[\lbol/L_{\odot}] - \log\left[10^{(\log[\lbol/L_{\odot}]-b)/a} + 1 \right],
\end{eqnarray}
where $a=0.69\pm0.03$ and $b=10.91\pm0.04$, as shown in
Figure~\ref{fig:bolvuv}.  The relationship between $\lbol$ and $\luv$
implies that galaxies with faint UV luminosities are less attenuated
than their UV-bright counterparts.  A recent study based on deep GOODS
and UDF ACS finds that UV-faint galaxies at $z\sim 2.5$ have
systematically bluer UV slopes ($\beta$) relative to UV-bright ones,
implying that the former may be less dusty \citep{bouwens09}.  Further, \citet{reddy08} and \citet{reddy09} argue based
on physical reasons that such UV-faint galaxies are unlikely to be as
attenuated on average as their UV-bright counterparts.  For example,
significant amounts of dust in these galaxies, combined with their
large number densities as inferred from the steep faint-end slope of
the UV LF, would result in stellar mass densities and an infrared
background significantly in excess of those measured \citep{reddy09}.
Taking this luminosity-dependent dust correction into account,
\citet{reddy08} and \citet{reddy09} demonstrate that such UV faint
galaxies dominate the bolometric luminosity density at $z\sim 2-3$.

Our present (larger) data set suggests a significant dependence
between bolometric and UV luminosity, as evidenced by the trend
between $24$~$\mu$m nondetection fraction and UV luminosity
(Figure~\ref{fig:luvundet}).  Specifically, this trend implies that
galaxies with faint $\luv$ are also on average less IR-luminous, in
turn suggesting that they are less bolometrically-luminous, than their
UV-bright counterparts.  The trend between $24$~$\mu$m nondetection
fraction and $\luv$ is further supported by a stacking analysis.
Specifically, we stacked the $24$~$\mu$m data in bins of $\luv$,
including both detections and nondetections, and limited the stack to
galaxies with $\lbol < 10^{12}$~L$_{\odot}$.  The latter restriction
is imposed because we are interested in determining whether a trend
exists between dustiness and UV luminosity for {\em typical} galaxies
at $z\sim 2$ (i.e., excluding bolometrically-luminous galaxies with
values of $\luv$ similar to those of galaxies with lower $\lbol$).
These stacked results yield average $\lbol$ consistent with the
prediction from combining the Meurer relation with the trend between
UV slope and continuum magnitude (e.g., \citealt{bouwens09}), as
illustrated in Figure~\ref{fig:bolvuv}.  Thus, our present sample
provides the first direct evidence (independent of the UV slope) for a
trend between UV luminosity on the one hand, and bolometric luminosity
and dust obscuration on the other.

At first glance, these results run counter to our previous analysis
that indicated no correlation between dustiness and UV magnitude for
UV-bright ($\rs < 25.5$) galaxies \citep{reddy08}.  More specifically,
in the previous analysis of the GOODS-North field, the stacked values
of $\lir/\luv$ as a function of UV magnitude had uncertainties that
were sufficiently large that we could not rule out the possibility of
no trend between dustiness and UV-magnitude (see Figure~11 of
\citealt{reddy08}; see also discussion in \S~\ref{sec:irxprev}).  Note
that despite this lack of trend observed with the smaller sample, we
investigated in detail the systematics introduced by assuming various
relations between dustiness and UV luminosity extending from UV-bright
galaxies (for which empircal constraints on the dust obscuration
factors exist) to UV-faint galaxies (see \citealt{reddy08, reddy09}).
The variation of IR and bolometric luminosity with UV luminosity
becomes more apparent with our larger sample
(Figures~\ref{fig:bolvuv},\ref{fig:luvundet}, \S~\ref{sec:irxprev}).

\subsubsection{Saturation of UV Luminosity}

The shallower-than-unity slope of the relationship between $\lbol$ and
dust obscuration implies that the UV luminosity turns over or
``saturates,'' at which point any additional star formation will be
optically-thick (Figure~\ref{fig:bolvuv}).  As we discuss below, this
saturation of the UV luminosity with increasing SFR likely explains
why no correlation was found previously between $\lbol$ and $\luv$
(e.g., as can also be seen by the scatter of the points in
Figure~\ref{fig:bolvuv}) since these spectroscopic samples probe a
relatively narrow range of UV luminosity around the value of
$L^{\ast}_{\rm UV}$.  The saturation point can only be ascertained
from the functional fit to the $\lbol$-$\lir/\luv$ relation and should
become more apparent with future measurements of the dust attenuation
of galaxies fainter than our spectroscopic limit.  In any case,
galaxies routinely found in {\em Spitzer} surveys at high redshift
(e.g., ULIRGs) are typically fainter in the UV than moderately
luminous galaxies found in optical surveys (e.g., LIRGs).  As noted
above, such ULIRGs appear to follow the same $\lbol$-dustiness
relation as UV-selected galaxies.  These results can be understood if
the dust obscuration, $\lir/\luv$, increases more rapidly than the
increase of {\em total} UV luminosity as the bolometric luminosity
increases.  This may largely explain why ULIRGs have IRX ratios far in
excess of the values predicted by their UV spectral slopes based on
the Meurer relation (\S~\ref{sec:meurer}, Figure~\ref{fig:bolcomp}).
The turnover in $\luv$ implies that even in the presence of galaxies
with very large SFRs at $z\sim 2$, the observed UV luminosity will
never be brighter than a certain value which, at $z\sim 2$,
corresponds to $\luv \approx 10^{11}$~L$_{\odot}$.

\begin{figure*}[!t]
\plottwo{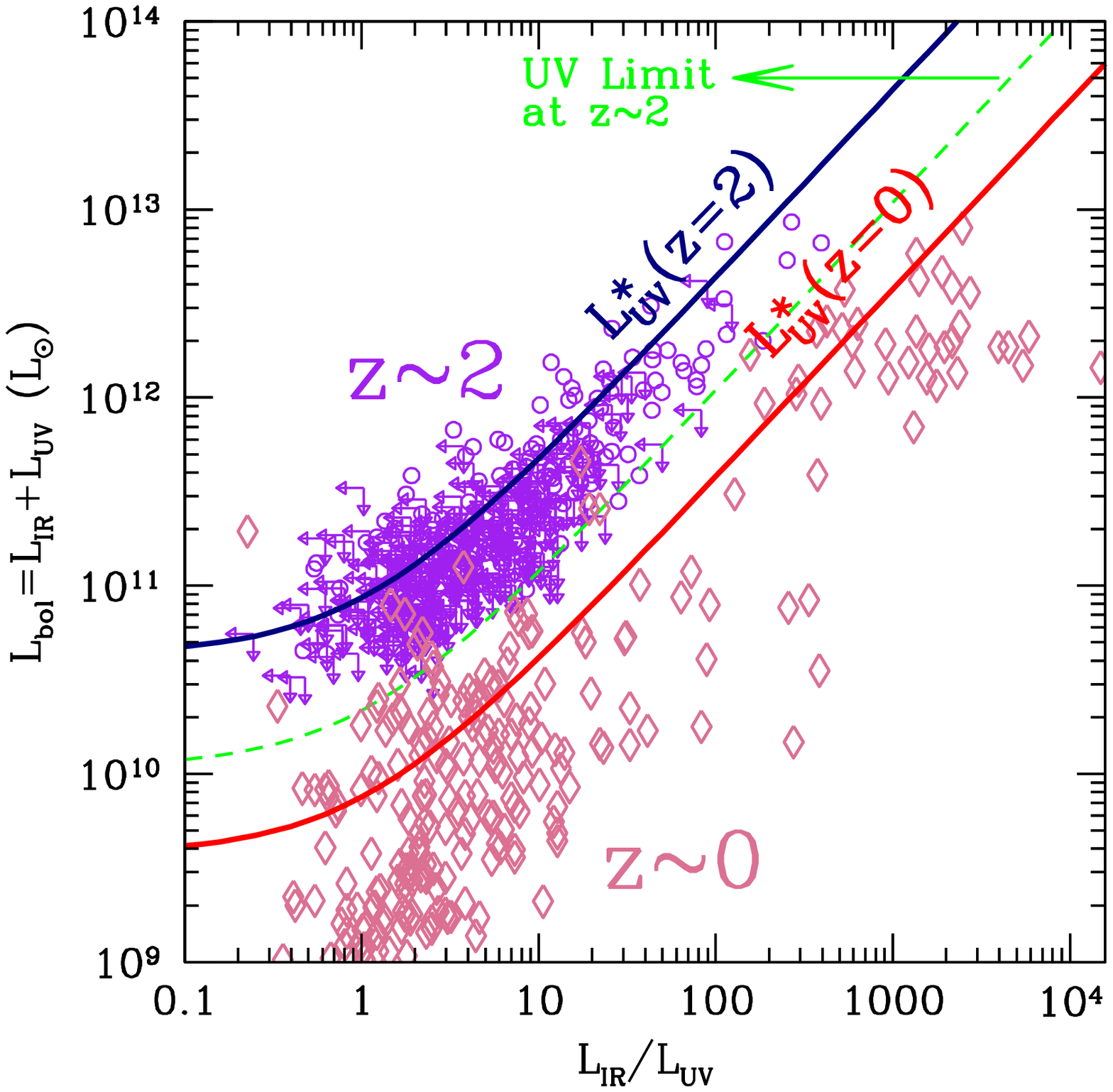}{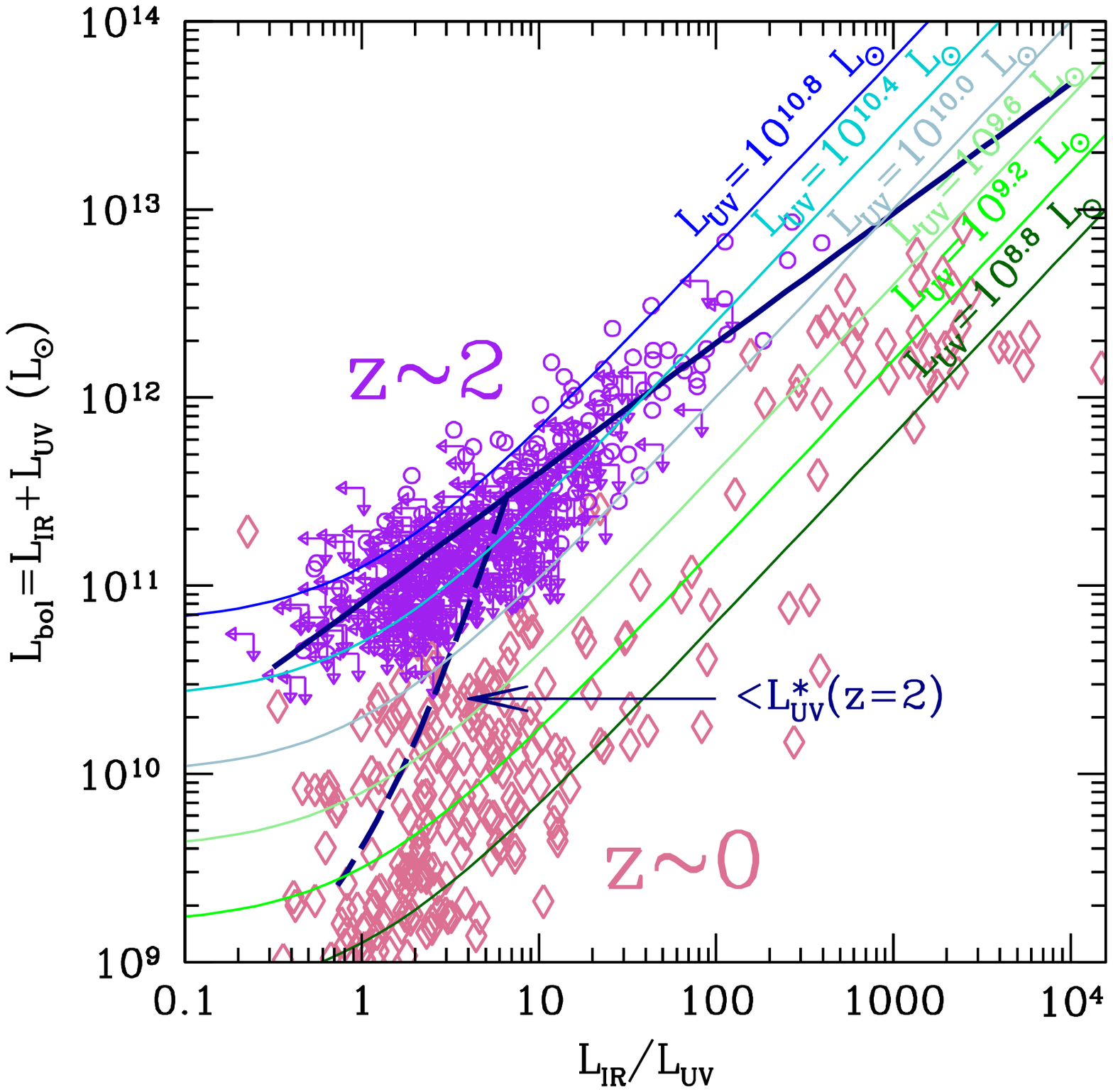}
\caption{({\em Left}): Bolometric luminosity versus dust attenuation
  for our $z\sim 2$ sample compared with that of local galaxies from
  \citet{bell03}, \citet{brandl06}, and \citet{huang09}.  Also shown
  are lines of constant UV luminosity at the value of $L^{\ast}_{\rm
    UV}$ at $z=2$ and $z=0$ (solid lines), and the UV spectroscopic
  limit at $z\sim 2$ (dashed line).  ({\em Right}): Same as left
  panel, in addition to the best-fit relation between bolometric
  luminosity and dust attenuation for $\luv\ga 10^{10}$~L$_\odot$ at
  $z\sim 2$ (thick solid line) and its extension to UV-faint galaxies
  (thick dashed line).  Also shown are lines of constant UV luminosity
  for $\luv = 10^{8.8} - 10^{10.8}$~L$_\odot$.}
\label{fig:luvbol}
\end{figure*}

\subsubsection{Implications for the Shape of the UV LF}

The connection between UV and bolometric luminosity has important
implications for the shape of the UV luminosity function.  In
particular, the saturation point for the observed UV luminosity occurs
approximately at the values of $L^{\ast}_{\rm UV}$ and $L^{\ast}_{\rm
  bol}$ at $z\sim 2$.  This is not surprising because, by definition,
$L^{\ast}$ is the point brighter than which the number density of
sources decreases exponentially.  This effect can also be seen by
comparing our previous determinations of the UV LF, which is
well-described by a Schechter function, with the bolometric LF, which
has a more power-law like shape and not as well defined of a ``knee''
\citep{reddy08, reddy09}.  The important point is that the saturation
effect is a direct result of the increasing extinction with $\lbol$.
Consequently, dust obscuration must be the dominant effect in
modulating the bright-end of the UV LF, and thus dictating the value
of $L^{\ast}$, at $z\sim 2$.

\subsection{Redshift Evolution}

Comparison with the relation derived locally (e.g., \citealt{buat06})
shows that $L^{\ast}$ galaxies at $z\sim 2$ exhibit $\lir/\luv$ ratios
that are a factor of 5 times lower than those of local galaxies with a
similar $\lbol$ \citep{reddy06a, reddy08}, implying a redshift
evolution in the extinction per unit star formation rate (or per unit
$\lbol$).  R06 suggested that this may reflect either an
evolution in the dust-to-gas ratios as galaxies age or a change in the
average sizes of the infrared and UV emitting regions of galaxies with
redshift.  In the following, we discuss the importance of the $\luv$
limit in accurately assessing this evolution, a consideration of the
$\lbol$-$\lir/\luv$ relation for UV-faint galaxies at high redshift,
and the implications of these results for the evolution of individual
galaxies.

\subsubsection{Dependence on $\luv$ Limit}

Quantifying properly the evolution in the $\lbol$-$\lir/\luv$ relation
requires a careful consideration of the $\luv$ limit used to compare
samples at different redshifts (see also the discussion in
\citealt{buat09}).  The strong redshift evolution in $L^{\ast}_{\rm
  UV}$ \citep{reddy09} implies that a relatively shallow UV luminosity
limit will exclude the parameter space in the $\lbol$-$\lir/\luv$
plane where most normal galaxies in the local universe lie
(Figure~\ref{fig:luvbol}), causing one to infer a milder redshift
evolution in the extinction per unit SFR.  It is clear that an
accurate comparison of these relations for typical galaxies requires a
$\luv$ limit that is sufficiently faint to detect these average
($L^{\ast}_{\rm UV}$) galaxies at all redshifts in question.  A
sufficiently faint $\luv$ limit reveals that $L^{\ast}_{\rm UV}$
galaxies at $z\sim 2$ are a factor of $\approx 10$ times less
attenuated than local galaxies of the same $\lbol$
(Figure~\ref{fig:luvbol}).

\subsubsection{Consideration of UV-faint Galaxies at High Redshift}

\begin{figure}[!t]
\plotone{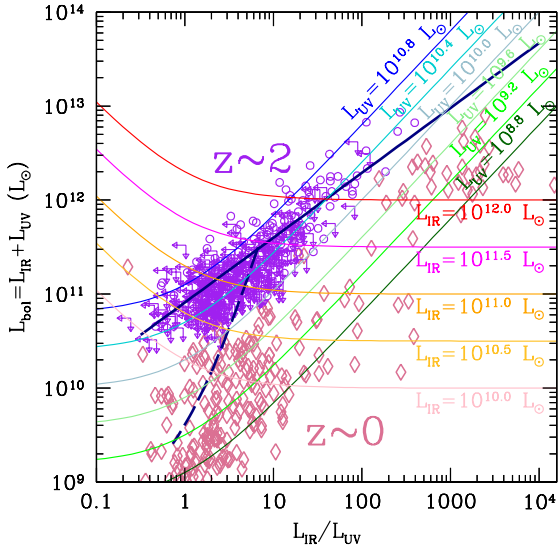}
\caption{Same as Figure~\ref{fig:luvbol} showing lines of constant
$\luv$ and $\lir$ luminosity.}
\label{fig:luvbol3}
\end{figure}

Clearly the $\luv$ limit is also an important consideration in our
interpretation of the $\lbol$-$\lir/\luv$ relation at $z\sim 2$.  This
limit is defined by the magnitude cut of our spectroscopic sample
($\rs=25.5$) and corresponds to $\luv\approx 10^{10}$~L$_{\odot}$
(Figures~\ref{fig:luvdist}, \ref{fig:luvbol}).  This limit is faint
enough to detect $L^{\ast}_{\rm UV}$ galaxies at $z\sim 2$, where
$L^{\ast}_{\rm UV}$ is determined not from the present sample alone,
but from a consideration of the UV luminosity function measured using
data extending $1.5$~mag fainter than our spectroscopic limit
\citep{reddy09}.  Therefore, the $\lbol$-$\lir/\luv$ relation derived
here is not just valid for the sample analyzed here, but valid for
$L^{\ast}_{\rm UV}(z=2)$ galaxies in general.  The saturation of
$\luv$ discussed above can also be seen in Figures~\ref{fig:luvbol}
and \ref{fig:luvbol3} where the relation between $\lbol$ and
$\lir/\luv$ at $z\sim 2$ crosses the same line of constant $\luv$ more
than once.

Let us now consider how the $\lbol$-$\lir/\luv$ relation may change
for galaxies fainter than the $\luv\approx 10^{10}$~L$_{\odot}$
threshold.  Our finding of lower bolometric luminosities for UV-faint
galaxies (\S~\ref{sec:bolvuv}) suggests that extending the
$\luv\approx 10^{10}$~L$_{\odot}$ limit to fainter luminosities is
unlikely to reveal a very large population of UV-faint galaxies with
$\lbol$ similar to those of UV-bright ones.  Therefore the
$\lbol$-$\lir/\luv$ relation is unlikely to be significantly broader
at $\lbol\sim 10^{11}-10^{12}$~L$_{\odot}$ than what we have measured
from the spectroscopic sample.

We have illustrated this point by using the relation between UV
luminosity and $\beta$ \citep{bouwens09} to infer the dust attenuation
and bolometric luminosities using the Meurer relation, thus extending
our measurement of the $\lbol$-$\lir/\luv$ relation to UV-faint
galaxies (Figure~\ref{fig:luvbol}).  Examined over a wider range in UV
luminosity, it becomes clear that the $\lbol$-$\lir/\luv$ relation at
$z\sim 2$ is defined by a sequence of galaxies with increasing $\lbol$
and $\lir/\luv$ with increasing UV luminosity.

\subsubsection{Evolution of Individual Galaxies}

Figure~\ref{fig:luvbol3} highlights another important issue.  While
the dustiness of galaxies of a given $\lbol$ increases with decreasing
redshift, this should not be miscontrued to suggest that a typical
$z\sim 2$ galaxy will evolve to become dustier but retain the same
$\lbol$.  Clustering analyses of $L^{\ast}$ galaxies at $z\sim 2$
imply that they evolve to reside in the bulges of spirals and low mass
ellipticals by $z\sim 0$ \citep{conroy08}.  Therefore, the
evolutionary track of any {\em single} galaxy in the
$\lbol$-$\lir/\luv$ plane is unlikely to be perpendicular to either
axis.  As galaxies on average fade between $z\sim 2$ and the present
day, reflecting the global decline of the star formation rate density
(e.g., \citealt{reddy09, reddy08, madau96, giavalisco96, steidel99}),
there will be a corresponding decrease in their $\luv$ and $\lir$, and
hence $\lbol$.  The critical point is that without taking into account
the offset between the $z\sim 2$ and $z\sim 0$ relations, one would
wrongly conclude that the dustiness observed in local $L^{\ast}_{\rm
  bol}$ galaxies is similar to that observed for galaxies of the same
bolometric luminosity at $z\sim 2$, when in fact the dustiness is
lower at a fixed $\lbol$ at higher redshift.

\begin{figure}[!t]
\plotone{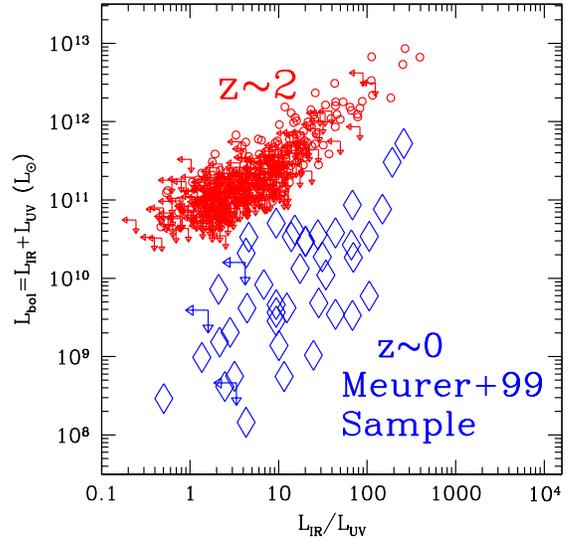}
\caption{Bolometric luminosity versus dust obscuration of the $z\sim
  2$ sample, compared to that of the local {\em IUE} sample that was
  used in large part to calibrate the \citet{meurer99} relation.}
\label{fig:iuecomp}
\end{figure}

The correlation between bolometric luminosity and dust attenuation
simply reflects the relationship between star formation rate and gas
surface density, i.e., the Schmidt law \citep{zoran06}.  An increase
in the dust-to-gas ratio with galaxy age would simply shift this
relationship with redshift, as is observed \citep{reddy06a, buat07,
  zoran06}.  In \S~\ref{sec:metals} we revisit the variation of
bolometric luminosity with dust attenuation in the context of the
oxygen abundances of starburst galaxies.  Geometrical effects are also
likely to play an important role in the offset between the $z\sim 2$
and $z\sim 0$ relations.  For instance, in Figure~\ref{fig:luvbol3},
many of the $z\sim 0$ galaxies with $\lbol\sim 10^{10}$~L$_\odot$ are
bulge-dominated spirals with star formation extending on scales of
several tens of kpc.  If most of the star formation occurs in the more
metal-poor parts of the galactic disks \citep{pilyugin04, zaritsky94,
  vila-costas92}, then the measured $\lir/\luv$ ratios would be
significantly lower than what we would infer based on the
globally-averaged dust-to-gas ratio, including the metal-rich
component associated with the bulge.

\subsubsection{Comparison with the Local {\em IUE} Sample}

The conclusion from the above observations is that some combination of
dust-to-gas ratio and size evolution will drive the observed
progression of the $\lbol$-$\lir/\luv$ relation with redshift.  A
consequence of this evolution is that we expect to see an offset in
the metallicity-luminosity relation with redshift, which we discuss in
\S~\ref{sec:metals}.  In practice, this redshift evolution in the
extinction per unit SFR implies that rest-UV selection allows one to
access galaxies with an increasingly larger range in $\lbol$ with
increasing redshift.  The increasing UV transparency with redshift was
noted by \citet{adel00, reddy06a} for UV-selected samples and shown to
be true for galaxies selected using other rest-optical and submm flux
criteria as discussed in \citet{reddy06a}.  \citet{daddi07a} also
discuss this issue in the context of the differences between local and
high-redshift ULIRGs.  The increasing dynamic range in bolometric
luminosity probed by UV selection is demonstrated more directly in
Figure~\ref{fig:iuecomp} where we compare typical galaxies at $z\sim
2$, for which the Meurer relation is found to be valid, with many of
the same galaxies that were used to calibrate the \citet{meurer99}
relation locally; the latter sample is drawn from observations with
the {\em International Ultraviolet Explorer} ({\em IUE};
\citealt{kinney93, heckman98, meurer99}).  Galaxies that were used to
calibrate the Meurer relation locally span the same range of IRX
($\lir/\luv$) as $L^{\ast}$ galaxies at $z\sim 2$, yet the local
sample spans luminosities that are anywhere from one to two decades
less luminous at a fixed $\lir/\luv$.  The Meurer relation appears to
hold for typical galaxies at $z\sim 2$ despite the fact that the
relation was calibrated on local galaxies that were significantly less
luminous.  This suggests that the UV SED alone can be used to recover
the total dust attenuation in galaxies with progressively larger
$\lbol$ at higher redshift, given that, at higher redshift, galaxies
are more transparent in the UV at a fixed $\lbol$
(Figures~\ref{fig:bolcomp}, \ref{fig:iuecomp}).

\subsection{Variations in $\lir$ and $\luv$ at a Given $\lbol$}

The contours of fixed $\luv$ and $\lir$ shown in
Figure~\ref{fig:luvbol3} emphasize the relative change in these
quantities as the dustiness is varied at a fixed $\lbol$ at $z\sim 2$.
At $\lbol \ga 2\times 10^{11}$~$L_{\odot}$, stochastic changes in the
dust obscuration (e.g., such as might be expected if the extinction is
patchy) results in $\luv$ to vary much more than $\lir$, simply
because a larger fraction of the total luminosity is obscured (e.g.,
see also \citealt{adel00}).  For faint galaxies with $\lbol \la
10^{10}$~L$_{\odot}$, an adjustment in the dust attenuation will do
little to alter the UV luminosity while $\lir$ changes more
dramatically.  Of course, none of these observations are particularly
surprising, given that as we adjust $\luv$ there must be a
corresponding change in $\lir$ to keep $\lbol$ constant.  

\subsection{Comparison with Previous Results}
\label{sec:irxprev}

Several times throughout the discussion above we have mentioned the
correlation between UV luminosity on the one hand, and bolometric
luminosity and dust attenuation on the other.  At face value, these
results contrast with previous investigations that have found no
correlation between UV luminosity and dustiness \citep{reddy06a,
  adel00}, as discussed briefly in \S~\ref{sec:bolfaint}.  Of course,
a constant average dust correction with UV luminosity naturally leads
to lower bolometric luminosities with decreasing UV luminosity.  It is
clear, however, that if UV-faint galaxies are significantly bluer than
their brighter counterparts \citep{reddy08, reddy09, bouwens09}, and
their bolometric luminosities are consistent with the Meurer
prediction (stacked points in Figure~\ref{fig:bolvuv}), then their
average dust obscuration must also be lower (dashed line in
Figure~\ref{fig:luvbol}).  Hence, their bolometric luminosities will
be correspondingly lower.  We find direct evidence of this from the
trend between $24$~$\mu$m detection fraction and UV luminosity
(Figure~\ref{fig:luvundet}), even for the present sample which is
larger but covers the same dynamic range in $\luv$ as previous studies
\citep{reddy06a, adel00}.  The trend between UV luminosity and UV
slope was found over a slightly larger dynamic range in $\luv$ than is
usually represented in spectroscopic samples \citep{bouwens09}.

The limited dynamic range and/or smaller samples may have contributed
to the apparent lack of correlation between UV luminosity and
dustiness observed before.  There are a couple of other reasons why
such a correlation may have been difficult to discern in previous
studies.  The first is that the saturation of $\luv$ occurs at a value
that lies in the range of $\luv$ that is typically probed in
spectroscopic surveys.  In other words, the non-monotonic behavior of
$\luv$ with bolometric luminosity for $\luv\ga 10^{11}$~L$_\odot$,
folded in with measurement errors, washes out the underlying trend
between $\luv$ and $\lbol$.  The trend between dustiness and UV
luminosity becomes more apparent once galaxies above the saturation
point are removed from the analysis (\S~\ref{sec:bolfaint}).  Second,
as seen in the previous section, the lack of correlation may be due to
the strong variance of $\luv$ with stochastic changes in the dust
obscuration for galaxies of moderate (or higher) luminosities (e.g.,
\citealt{adel00}).  It is clear that with larger samples, superior
(direct) tracers of dust emission, and a larger dynamic range, one can
obtain a more complete picture of how dust obscuration varies with
other galaxy properties.

\subsection{Summary}

In this section we have considered the functional form of the
correlation between bolometric luminosity and dust obscuration, its
consequence for the shape of the UV luminosity function at $z\sim 2$,
and its redshift evolution.  For galaxies brighter than our
spectroscopic limit of $\luv \approx 10^{10}$~L$_{\odot}$, we find a
tight positive correlation between bolometric luminosity and dust
obscuration ($\lbol$-$\lir/\luv$).  Our sample provides the first
direct evidence for the decrease of infrared luminosity with
decreasing UV luminosity.  Our results suggest that dust obscuration
is likely the dominant effect in modulating the bright-end of the UV
LF as evidenced by the saturation of UV luminosity with increasing
star formation rate.  We demonstrate the importance of the $\luv$
limit in quantifying the redshift evolution of the $\lbol$-$\lir/\luv$
relation.  Using a sufficiently faint limit, we find that $L^{\ast}$
galaxies at $z\sim 2$ have dust obscuration ratios ($\lir/\luv$) that
are roughly a factor of $10$ times lower than those of local galaxies
of the same bolometric luminosity.  We examine this offset in the
context of metallicity in the next section.

\section{Relationship between Dust Attenuation and Metallicity at $z\sim 2$}
\label{sec:metals}

It is clear from the discussion above that an increase in the
dust-to-gas ratios with age plays an important role in the observed
evolution of the correlation between bolometric luminosity and dust
attenuation.  More generally, because most of the metals in the ISM
will be contained in (or depleted onto) the same dust grains that
attenuate starlight, we expect that the dust attenuation should
correlate directly with metallicity and hence stellar mass.  The
near-IR spectroscopic data for LBGs at $z\sim 2$ have been used to
constrain their gas-phase metallicities using the [NII]/H$\alpha$
ratio.  \citet{erb06a} compute the stellar mass-metallicity relation
at $z\sim 2$ in this manner and find that the relation is offset from
the local one \citep{tremonti04} such that at a fixed stellar mass
galaxies at $z\sim 2$ are $\approx 2$ times less metal-rich than
present-day galaxies.  This offset likely reflects the larger gas
fractions at a given stellar mass with increasing redshift (see
discussion in \citealt{erb06a}).

\begin{figure*}[!t]
\plottwo{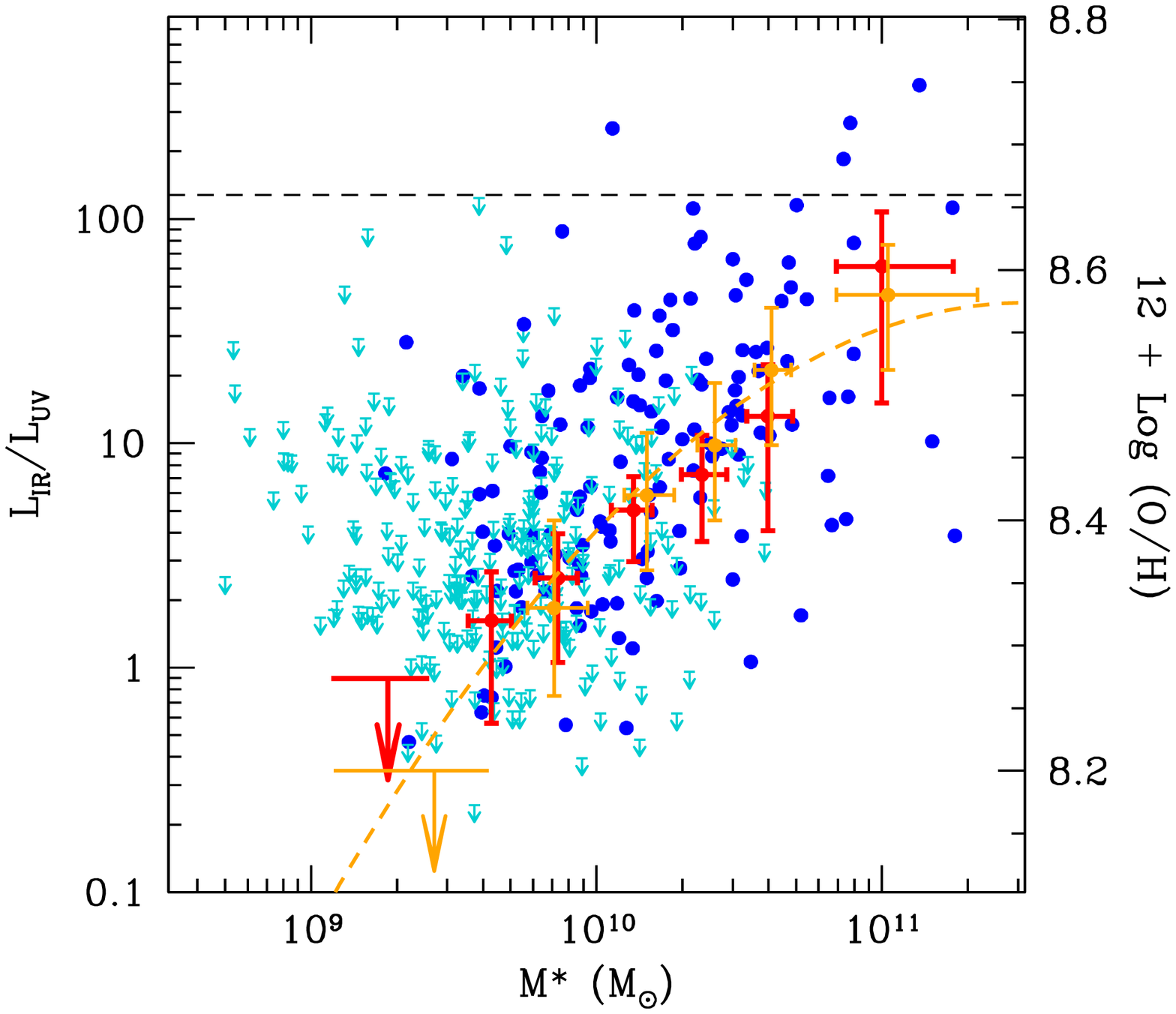}{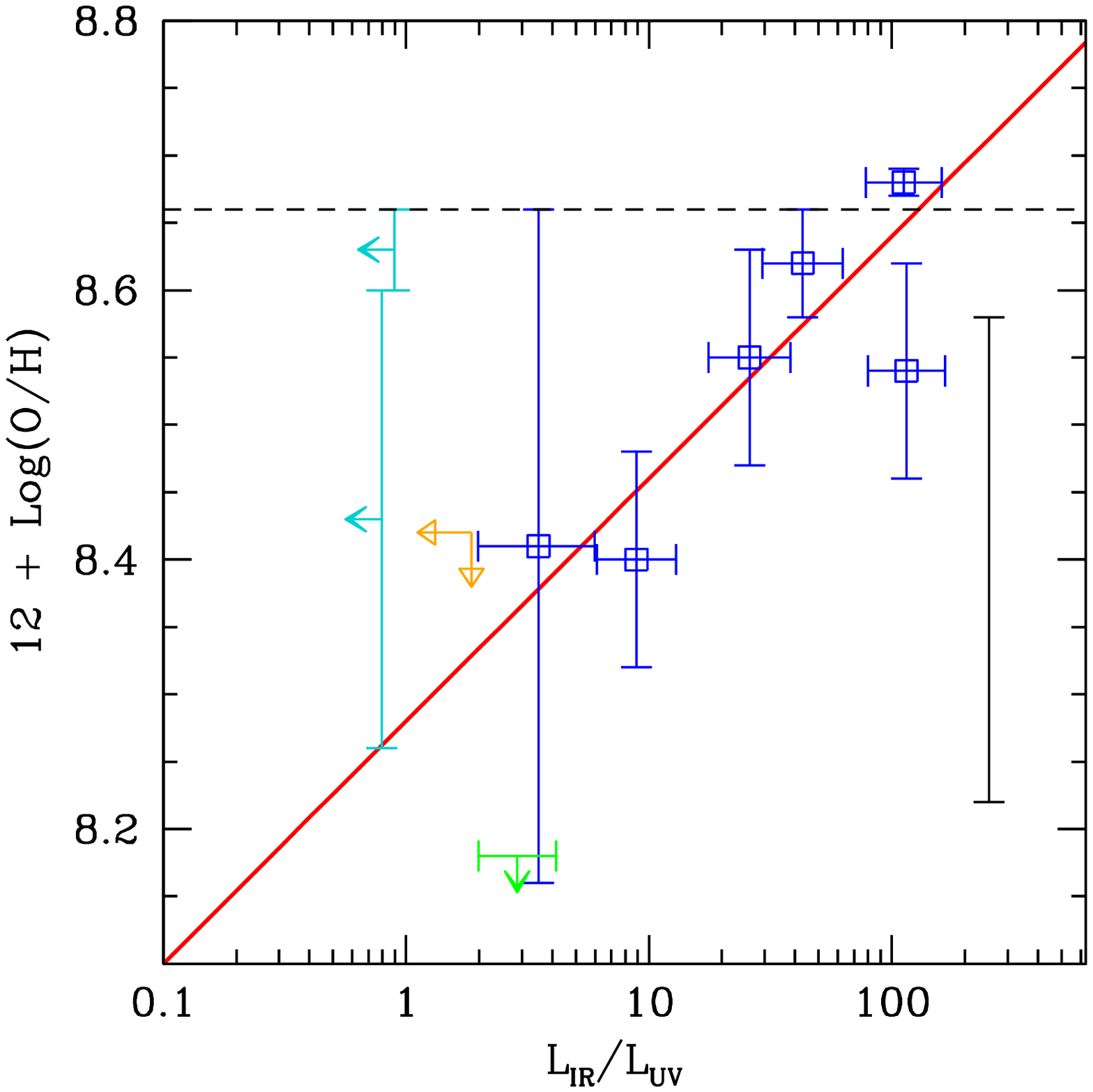}
\caption{({\em Left}): Dust attenuation, parameterized as $\lir/\luv$,
  as a function of stellar mass for 392 galaxies observed with MIPS.
  Galaxies detected and undetected at $24$~$\mu$m are indicated by the
  circles and downward pointing arrows, respectively.  The average
  stacked values of the attenuation are indicated by the red points.
  The mass-metallicity relation from \citet{erb06a} is denoted by the
  orange points with the metallity scale indicated on the right-hand
  axis.  The mass-metallicity relation is scaled to match the stacked
  estimates for the attenuation according to Eq.~\ref{eq:attmet}.  The
  dashed horizontal line indicates solar metallicity
  \citep{asplund04}.  ({\em Right}): Measurements of attenuation and
  oxygen abundance (derived from the N2 index; \citet{pettini04}) for
  individual star-forming galaxies at $z\sim 2$.  Points are color
  coded to reflect whether they have upper limits in attenuation
  and/or metallicity.  The vertical black error bar denotes the random
  scatter between the [NII]/H$\alpha$ ratio and oxygen abundance.  The
  solid red line indicates the relation found by normalizing the
  mass-attenuation relation to match the mass-metallicity relation
  ({\em left} panel).  The dashed horizontal line indicates solar
  metallicity \citep{asplund04}.}
\label{fig:metals}
\end{figure*}

\subsection{Correlation at $z\sim 2$}

To test our expectation of a correlation between metallicity and dust
attenuation, we have investigated the variation of $\lir/\luv$ with
stellar mass for the sample of 392 galaxies (Figure~\ref{fig:metals}).
Assuming that the $24$~$\mu$m undetected galaxies have a similar
scatter in $\lir/\luv$ at a given mass as the $24$~$\mu$m detected
galaxies, we find a scatter between attenuation and stellar mass of
$\approx 0.46$~dex about a linear fit between the two.  The trend
between attenuation and mass can be visualized more easily by
examining the mean attenuation in bins of stellar mass as inferred
from a stacking analysis.  To compute the average $\lir/\luv$, we
first stacked the 24~$\mu$m emission for galaxies in each bin of
stellar mass to find their average $\lir$ luminosity.  The average
$\lir$ luminosity is then combined with the average $\luv$ luminosity
for galaxies in each bin of stellar mass to compute the $\lir/\luv$
ratio.  The error in the ratio takes into account the error in the
stacked estimate of the $\lir$ luminosity and the error in the mean of
the $\luv$ luminosity.  Doing this, we find that galaxies with stellar
masses $\ga 10^{11}$~M$_{\odot}$ are roughly 100 times more attenuated
on average than those with stellar masses $\la 10^{9.5}$~M$_{\odot}$.
Because the gas-phase metallicity also appears to be significantly
correlated with stellar mass (and in the same direction as that
observed for the mass-attenuation relation), it suggests a close
connection between dust attenuation and gas-phase metallicity, in
accord with expectations.  Relating the mean obscuration in a given
bin of stellar mass with the metallicity expected for that stellar
mass from the mass-metallicity relation, we obtain the following
empirical relation between obscuration and gas-phase metallicity as
measured via the oxygen abundance:
\begin{eqnarray}
12 + \log (O/H) = \nonumber \\
(0.18\pm0.03) \log(\lir/\luv) + (8.28\pm0.03) \nonumber \\
{\rm for}~1\la \lir/\luv \la 40.
\label{eq:attmet}
\end{eqnarray}
The upper limit to which the calibration is valid ($\lir/\luv \approx
40$) corresponds to the point at which [NII]/H$\alpha$ begins to
saturate for galaxies with close to solar metallicity (see next
section).  This conversion allows one to estimate to within $0.54$~dex
random scatter the gas-phase metallicity expected for a given dust
attenuation for $L^{\ast}$ galaxies at $z\sim 2$.  This scatter
includes both the scatter in the mass-metallicity relation
\citep{erb06a} and that in the relation between the [NII]/H$\alpha$
ratio and oxygen abundance.  This relation can be useful in practice
because measuring the dust obscuration for large numbers of individual
galaxies is observationally more feasible than measuring gas-phase
metallicities; the latter requires near-IR spectroscopy at these
redshifts.  Nonetheless, we caution against the use of this relation
for dust obscurations that are dissimilar to the ones used in the
calibration, given the very limited dynamic range in metallicity
probed by the current sample.  Figure~\ref{fig:metals} shows the
comparison between dust attenuation, stellar mass, and metallicity.

Recall that Eq.~\ref{eq:attmet} is derived from the average
attenuation and metallicity of galaxies with a given stellar mass.  To
better gauge how well this relationship holds for individual galaxies,
we examined sources in our sample that also have published individual
metallicity determinations (see Table~9 of \citealt{law09}).  To
remove any potential systematics between metallicity calibrations, we
considered only galaxies where oxygen abudances were inferred from the
N2 index \citep{pettini04}.  While there were only 10 galaxies in our
sample that met these requirements, their dust attenuations and
metallicities are consistent with our conversion between the two
within the measurement errors ($\approx 0.12$~dex scatter between the
individual measurements and the relationship determined from stacking;
Figure~\ref{fig:metals}).  Obviously, larger samples of galaxies with
individual metallicity measurements, such as those made possible with
multi-object near-IR spectrographs (e.g., MOSFIRE;
\citealt{mclean08}), will enable more detailed investigations of the
connection between dust attenuation and gas-phase metallicity.

\begin{figure*}[!t]
\plottwo{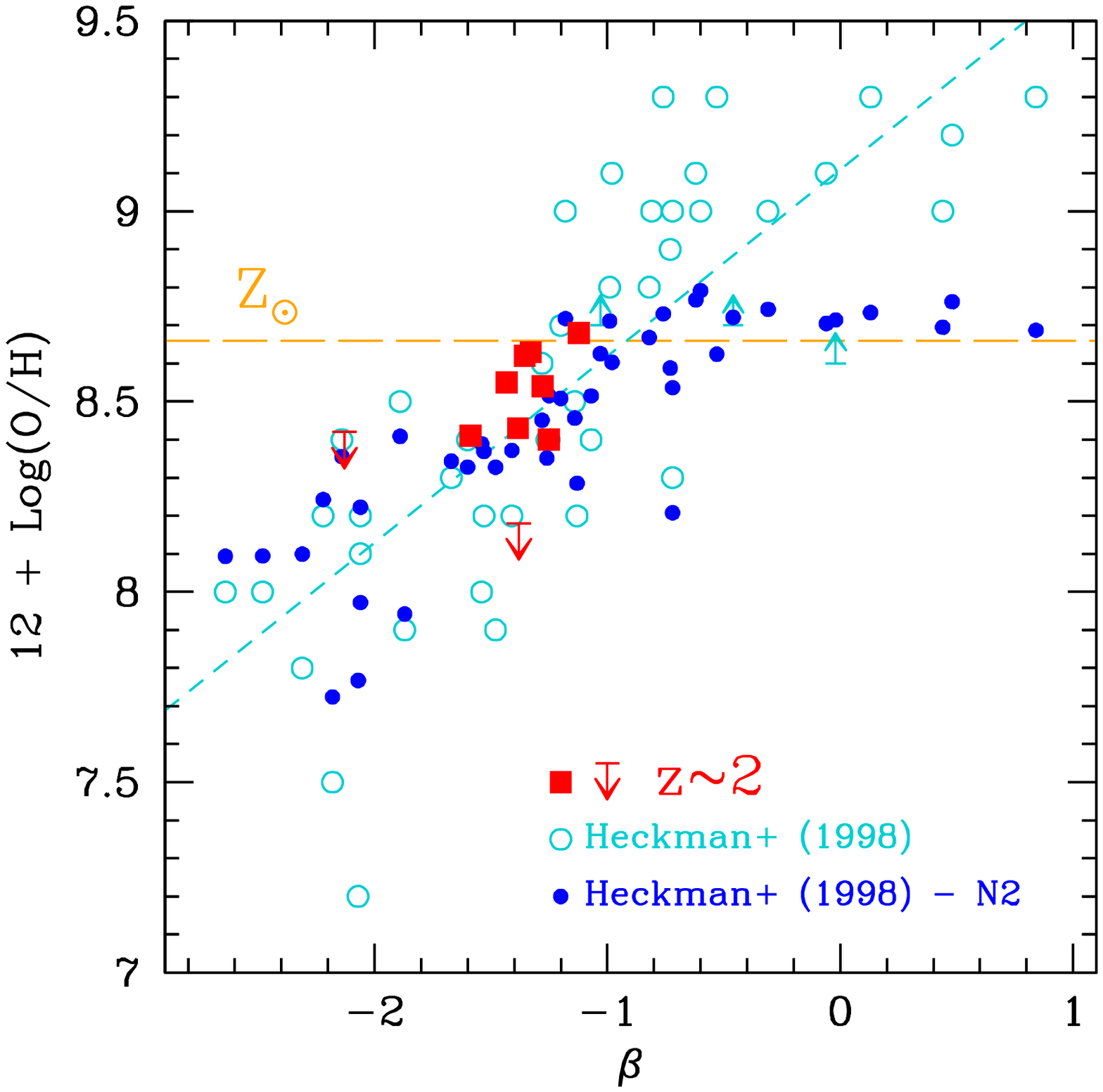}{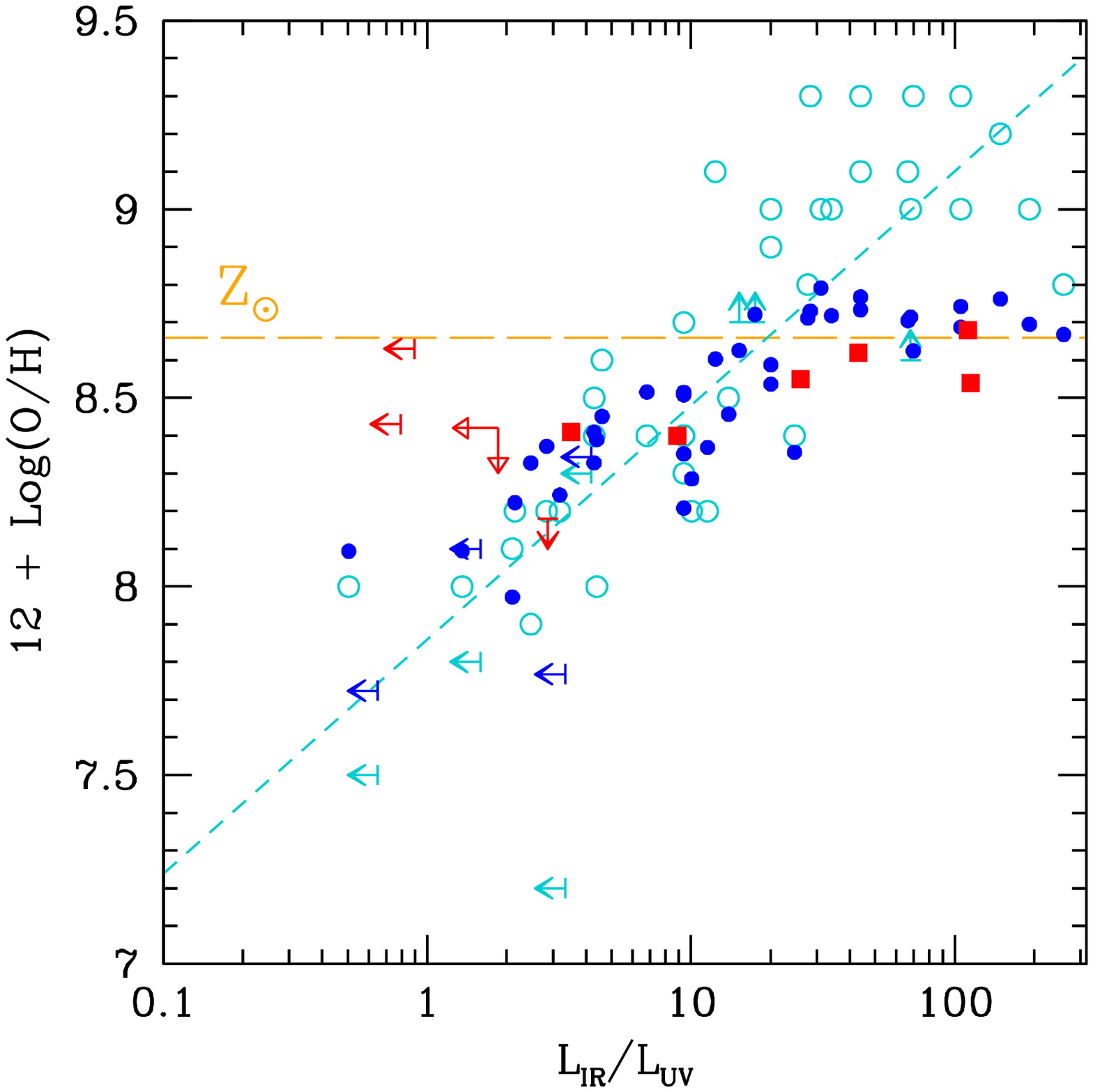}
\caption{({\em Left}): Oxygen abundance versus UV slope, $\beta$, for
  galaxies at $z\sim 2$.  For comparison, data for local starburst
  galaxies from \citet{heckman98} are provided, both using the
  original metallicity calibrations of that study ({\em open circles})
  and updating their results for the [NII]/H$\alpha$-derived
  metallicity ({\em filled circles}).  Their best-fit linear relation
  (for the original data) is indicated by the short-dashed line.  The
  long-dashed line denotes solar metallicity \citep{asplund04}.  ({\em
    Right}): Same as left panel for oxygen abundance versus dust
  obscuration, $\lir/\luv$.}
\label{fig:metloc}
\end{figure*}

\subsection{Comparison with the Local Relations}

We can gain a broader perspective on the $z\sim 2$ results by
comparing them to the local trend between metallicity and dust
obscuration, where locally the galaxy metallicities are more robustly
measured using a variety of calibrations and observations cover a
larger dynamic range in metallicity.  \citet{heckman98} find strong
positive trends of oxygen abundance (O/H) with $\lir/\luv$ and $\beta$
for a sample of local starburst galaxies (Figure~\ref{fig:metloc}).
They interpret these trends as a reflection of the increasing
extinction of UV emission (and reddening of the UV slope) as the
dust-to-gas ratio increases with gas-phase metallicity.

Before comparing with the $z\sim 2$ results, we must first account for
the systematic differences in the way that oxygen abundances are
inferred locally versus at high redshift \citep{kewley08}.  As
discussed above, the abundances for the $z\sim 2$ sample are derived
based on the calibration of [NII]/H$\alpha$ with O/H
\citep{pettini04}.  Alternatively, the abundances tabulated in
\citet{heckman98} are based primarily on either direct
temperature-sensitive ($T_{\rm e}$) methods or the R23 calibration.
For a robust comparison with the $z\sim 2$ sample, we searched the
literature to find [NII]/H$\alpha$ measurements for all of the
\citet{heckman98} galaxies.  The comparison between the metallicities
tabulated in \citet{heckman98} and those derived from the N2 index is
shown in Figure~\ref{fig:metloc}.  The two sets of metallicity
measurements agree well up to solar metallicity, at which point the
[NII]/H$\alpha$ indicator saturates due to nitrogen becoming the
dominant coolant, an effect that has been noted elsewhere
\citep{pettini04, kewley02, baldwin81}.

Having remeasured the metallicities for the local sample using the
same abundance indicator adopted at high redshift, we find that the
variations in O/H with $\beta$ and IRX for $L^{\ast}$ galaxies at
$z\sim 2$ are in general agreement with the local relations
(Figure~\ref{fig:metloc}).  The implication of this agreement is that
for a fixed metallicity, the $z\sim 2$ and $z\sim 0$ galaxies have similar
dust-to-gas ratios, despite the fact that the high-redshift ones are $10-20$
times more bolometrically luminous.  It is here that we see another
manifestation of the redshift evolution of the correlation between
bolometric luminosity and dustiness.  As discussed in
\S~\ref{sec:bol}, $z\sim 2$ galaxies are significantly more luminous
at a fixed dust obscuration than local galaxies.  We see a similar
effect in the context of metallicity, in the sense that galaxies with
a fixed metallicity (or dust-to-gas ratio) are significantly more
luminous at high redshift.  This result is a natural expectation given
the positive correlation between dust attenuation and metallicity
(Figure~\ref{fig:metloc}).

\subsection{Bolometric Luminosity - Metallicity Relation at $z\sim 2$}

The progression of metallicity with increasing obscuration
(Figures~\ref{fig:metals}) and the strong correlation between
bolometric luminosity and dust attenuation (\S~\ref{sec:bol}) imply
the existence of a luminosity-metallicity relation at $z\sim 2$ (e.g.,
see also \citealt{erb06a, shapley04}).  Further, the similarity in
dust-to-gas ratio and difference in luminosity suggests a significant
redshift evolution in the optical luminosity - metallicity relation,
as has been noted by \citet{erb06a, shapley04} in the context of
$z\sim 2$ galaxies, and noted elsewhere for samples primarily at lower
redshift (e.g., \citealt{kobulnicky04, kobulnicky03, maier06, maier04,
  salzer05, lara09}).  Quantitatively, a LIRG with $\lbol \sim
10^{11}$~L$_{\odot}$ at $z\sim 2$ is on average a third less
metal-enriched than a galaxy with the same (LIRG) bolometric
luminosity locally.  Similarly, the results suggest that high-redshift
galaxies are scaled up versions of local starbursts where, for a fixed
metallicity, the high-redshift ones are significantly more luminous
than the local ones.

These observations can be understood if high-redshift galaxies are on
average less-evolved than local galaxies, where the high-redshift ones
have larger gas fractions and are forming stars in a less metal-rich
environment (e.g., \citealt{erb06a, reddy06a}).  The redshift
evolution of both the luminosity-metallicity and the
luminosity-obscuration relations is then largely reflective of the
underlying chemical evolution of galaxies as they age.  As we have
seen (\S~\ref{sec:bol}), this evolution has important implications for
the relative transparency (or ``optical-depth'') of galaxies and their
amenability to UV-selection at ever-increasing redshifts.

\section{CONCLUSIONS}
\label{sec:conclusions}

We use ground-based UV imaging, near-IR spectroscopy, and {\em
  Spitzer} MIPS $24$~$\mu$m imaging of a large sample of Lyman Break
galaxies to investigate how the 8~$\mu$m luminosity ($\mirlum$) is
related to the H$\alpha$ luminosity ($\lha$), infrared luminosity
($\lir$), and star formation rate of $L^{\ast}$ galaxies at $z\sim 2$.
{\em Chandra} X-ray data are used to provide an independent check of
the average star formation rates derived in this manner.  We then use
the derived relationships to estimate the bolometric luminosities
($\lbol$) and dust attenuation of typical galaxies at $z\sim 2$.  Our
main conclusions are as follows:

1.  Using a sample of 90 Lyman Break Galaxies with H$\alpha$
spectroscopic and narrowband observations and MIPS $24$~$\mu$m
imaging, we find a tight correlation between $\mirlum$ and $\lha$ with
0.24 dex scatter.  We combine this result with the Kennicutt
relations, taking care to account for the unobscured component of the
star formation rate, to derive relations between $\mirlum$ and
$\lir$/SFR.

2.  Based on a larger sample of 392 galaxies with MIPS observations,
we find that the rest-frame UV slopes ($\beta$) of typical
star-forming galaxies at $z\sim 2$ with ages $\ga 100$~Myr correlate
significantly with dust attenuation, parameterized by the ratio of
infrared-to-UV luminosity, $\lir/\luv$.  Galaxies with flatter (bluer)
$\beta$ are less dusty on average than those with redder $\beta$.
Further, the correlation between $\beta$ and dust attenuation is
indistinguishable from that established for local UV-starburst
galaxies \citep{meurer99, calzetti00}, the latter of which is almost
always used to infer the dust attenuation of UV-selected galaxies at
high redshift.  We demonstrate here that the local correlation can be
used to infer the extinction and bolometric luminosities of
$10^{10}\la \lir \la 10^{12}$~L$_{\odot}$ galaxies at $z\sim 2$ to
within a scatter of $0.4$~dex.  Galaxies with the largest bolometric
luminosities of $\lbol \ga 10^{12}$~L$_{\odot}$ have bluer $\beta$
than their dust attenuations would imply based on the local
correlation, or the correlation for the vast majority of star-forming
galaxies at $z\sim 2$.  This effect is likely due to the fact the
majority of the star formation in these bolometrically luminous and
heavily attenuated galaxies is optically-thick at UV wavelengths (see
also point 5 below).

3.  Separately, $\la 13\%$ of our $z\sim 2$ sample consists of young
galaxies with inferred ages of $\la 100$~Myr.  Unlike their older and
more typical counterparts, these young galaxies are significantly less
attenuated at a given $\beta$, as evidenced by their larger
non-detection rate at $24$~$\mu$m and their non-detections in the
stacked $24$~$\mu$m and X-ray images.  These observations suggest that
young galaxies may follow an extinction curve that is different than
the usually assumed Meurer/Calzetti; the data for the young galaxies
are consistent with an SMC-like extinction curve.  If this is the
case, then their dust obscuration may be up to a factor of 2-3 lower
than the values obtained by assuming the local correlation between
$\beta$ and dust attenuation.

4.  We verify our previous result that galaxies with larger bolometric
luminosities are more heavily attenuated by dust, and galaxies with
redder $\beta$ are also more bolometrically luminous on average than
those with bluer $\beta$.  Comparison with the local correlation
between $\lbol$ and dust attenuation implies that $L^{\ast}$ galaxies
at $z\sim 2$ with $\lbol \sim 10^{11}$~L$_{\odot}$ are an order of
magnitude less dusty than local galaxies with a similar $\lbol$.  Such
an effect may be related to an increase of dust-to-gas ratio as
galaxies age.  The redshift evolution in the extinction per unit SFR
implies that in practice UV selection gives access to galaxies with an
increasingly larger range in $\lbol$ at increasing redshift.  It also
implies that the UV SED alone can be used to recover the total dust
attenuation in galaxies with progressively larger $\lbol$ at higher
redshift, given that, at higher redshift, galaxies are more
transparent in the UV at a fixed $\lbol$.

5.  We recast the correlation between bolometric luminosity and dust
attenuation at $z\sim 2$ in terms of observed UV luminosity to come to
the following conclusions.  First, galaxies with faint UV luminosities
are expected to be less attenuated than their UV-bright counterparts.
This result is supported by our data that indicate that UV-faint
galaxies are less IR luminous (and less bolometrically luminous) than
UV-bright ones.  Second, we observe that galaxies with very large star
formation rates (e.g., ULIRGs) are less UV luminous per unit SFR than
galaxies with lower SFRs (e.g., LIRGs), implying that the dust
covering fraction is likely increasing more rapidly with SFR than the
intrinsic UV luminosity.  This effect results in a saturation of the
UV luminosity.  This saturation occurs at the value of $L^{\ast}$ at
these redshifts, implying that the bright-end of the UV LF at $z\sim
2$ is likely modulated by dust obscuration.

6.  Motivated by the expectation of a direct correspondence between
extinction and metallicity, we have examined the relationship between
dust attenuation and stellar mass.  Galaxies with stellar mass $\ga
10^{11}$~M$_{\odot}$ are almost 100 times more dusty on average than
those with masses $\la 10^{9.5}$~M$_{\odot}$.  The monotonically
increasing mass-attenuation and mass-metallicity relations imply a
close connection between attenuation and metallicity, allowing us to
provide an empirical calibration between the two.  This calibration is
verified with a small sample of $z\sim 2$ galaxies with individual
metallicity determinations.  Comparison with the local relationship
between metal abundance and dust obscuration suggests that $L^{\ast}$
galaxies at $z\sim 2$ have a similar ratio of dust-to-metals as local
starbursts, despite the high-redshift galaxies' bolometric
luminosities being a factor of 10 to 20 times larger than local
galaxies.  Our results imply a redshift evolution in the
luminosity-metallicity and luminosity-obscuration relations that
reflects the underlying chemical evolution of galaxies as they age.

We have utilized all the data at our disposal, including UV imaging,
H$\alpha$ narrowband and spectroscopic observations, and {\em Spitzer}
MIPS $24$~$\mu$m imaging, to investigate the viability of rest-frame
$8$~$\mu$m luminosity as a star formation rate and dust indicator.  We
go on to show how these data indicate that the UV slopes of
high-redshift galaxies are sensitive to dust in the same way as they
are in the local universe; thus the UV slope is an important proxy for
inferring the bolometric luminosities of high-redshift galaxies in the
absence of longer wavelength data.  Further analysis will benefit from
observations with near-IR multi-object spectrographs allowing
simultaneous coverage of at least two near-IR bands, which will allow
for direct measurements of the nebular reddening via the Balmer
decrement for $z\sim 2$ galaxies.  Future observations with the {\em
  Herschel} Space Telescope will no doubt greatly extend our results.
These new observations will directly sample the longer wavelength
(rest-frame $30$~$\mu$m) dust emission of at least luminous LIRGs at
$z\sim 2$, and will yield more robust measurements of the bolometric
luminosities of more typical galaxies at $z\sim 2$.

\acknowledgements

We thank the staff of the Keck and Palomar Observatories for their
help in obtaining the data presented here.  Support for N. A. R. was
provided by NASA through Hubble Fellowship grant HST-HF-01223.01
awarded by the Space Telescope Science Institute, which is operated by
the Association of Universities for Research in Astronomy, Inc., for
NASA, under contract NAS 5-26555.  Additional support has been
provided by research funding for the {\em Spitzer} Space Telescope
Legacy Science Program, provided by NASA through contracts 1224666 and
1287778, issued by the Jet Propulsion Laboratory, California Institute
of Technology.  C. C. S. has been supported by grants AST 03-07263 and
AST 06-06912 from the National Science Foundation and by the David and
Lucile Packard Foundation.

\end{document}